\documentclass[final,3p,times,twocolumn]{elsarticle}
\usepackage{amssymb}
\usepackage{amsmath}
\usepackage{systeme}
\usepackage{amsmath}
\usepackage{booktabs}
\usepackage{amssymb}
\usepackage{multirow}
\usepackage{tabularx}
\usepackage{subfiles}
\usepackage[caption=false,font=footnotesize,labelfont=sf,textfont=sf]{subfig}

\journal{Sustainable Energy, Grids and Networks}

\begin{document}

\begin{frontmatter}


 %\title{Title\tnoteref{label1}}
 %\tnotetext[label1]{}

\title{Data-driven Small-signal Modeling for Converter-based Power Systems}
 
 \author[inst]{Francesca Rossi\corref{cor1}}%\fntref[label2]}
 \ead{francesca.rossi@upc.edu}
 \author[inst]{Eduardo Prieto-Araujo}%\fntref[label3]}
 \author[inst]{Marc Cheah-Mane}
 \author[inst]{Oriol Gomis-Bellmunt}
 
%\fntext[label2]{CITCEA-UPC, Department of Electrical Engineering, Universitat Politècnica de Catalunya, Av. Diagonal 647, 08028 Barcelona, Spain}
\cortext[cor1]{Corresponding author at: Centre d’Innovació Tecnològica en Convertidors Estàtics i Accionaments (CITCEA-UPC), Departament d’Enginyeria Elèctrica, Universitat Politècnica de Catalunya, ETS d’Enginyeria Industrial de Barcelona, Av. Diagonal, 647, 2nd floor, 08028 Barcelona, Spain}
\affiliation[inst]{organization={CITCEA-UPC, Department of Electrical Engineering, Universitat Politècnica de Catalunya},
            addressline={Av. Diagonal 647},
             city={Barcelona},
             postcode={08028},
             state={Spain},
             country={Spain}}

 %%\fntext[label3]{CITCEA-UPC, Department of Electrical Engineering, Universitat Politècnica de Catalunya, Av. Diagonal 647, 08028 Barcelona, Spain}


%\author[inst1]{Author One}

%\affiliation[inst1]{organization={Department One},%Department and Organization
%            addressline={Address One}, 
%            city={City One},
%            postcode={00000}, 
%            state={State One},
%            country={Country One}}

%\author[inst2]{Author Two}
%\author[inst1,inst2]{Author Three}

%\affiliation[inst2]{organization={Department Two},%Department and Organization
%            addressline={Address Two}, 
%            city={City Two},
%            postcode={22222}, 
%            state={State Two},
%            country={Country Two}}

\begin{abstract}
This article details a complete procedure to derive a data-driven small-signal-based model useful to perform converter-based power system related studies. To compute the model, Decision Tree (DT) regression, both using single DT and ensemble DT, and Spline regression have been employed and their performances have been compared, in terms of accuracy, training and computing time. The methodology includes a comprehensive step-by-step procedure to develop the model: data generation by conventional simulation and mathematical models, databases (DBs) arrangement, regression training and testing, realizing prediction for new instances. The methodology has been developed using an essential network and then tested on a more complex system, to show the validity and usefulness of the suggested approach. Both power systems test cases have the essential characteristics of converter-based power systems, simulating high penetration of converter interfaced generation and the presence of HVDC links. Moreover, it is proposed how to represent in a visual manner the results of the small-signal stability analysis for a wide range of system operating conditions, exploiting DT regressions. Finally, the possible applications of the model are discussed, highlighting the potential of the developed model in further power system small-signal related studies.
\end{abstract}



\begin{keyword}
Data-driven modeling \sep Decision Tree regression \sep Spline regression \sep Small-signal stability \sep Participation Factors \sep Power systems.
\end{keyword}

\end{frontmatter}


\section{Introduction}
The use of data-driven techniques is proliferating in the field of power system analysis \cite{vom2020data} thanks to the large amount of data that is being captured by network operators \cite{entso2015tso,hirth2018entso,hirth2020open}. An intelligent and efficient use of this data by means of advanced data-based techniques can provide relevant information to improve the network planning, operation and control \cite{entso2015tso}. However, the access to real data might not always be possible, due to Intellectual Property (IP) restrictions \cite{hirth2020open} or lack of measurements \cite{idehen2019pmu}, and provides knowledge of a limited set of scenarios, i.e. power system operating conditions.
Therefore, the use of high-fidelity simulations and mathematical models for the construction of data-driven models can be extremely useful for system studies. This is specially important if extreme scenarios, which can endanger the system stability, are to be considered. In these cases, data-driven models based on simulation can be the only choice as there is no available data.

Assuming that the simulation and mathematical models are correct and have been validated, they allow generating representative system response data for several different scenarios. This is independent from whether the scenarios are likely to occur or have been experienced in real operation. Engineers have complete freedom to select the variables that are measured and used for system response estimation. Once the data set has been generated in an adequate manner, data-driven techniques can be used to extract additional information from the simulation and mathematical models, that conventional engineering-based techniques might not be able to provide. 

\begin{figure}[!ht]
    \centering
  \includegraphics[width=\linewidth]{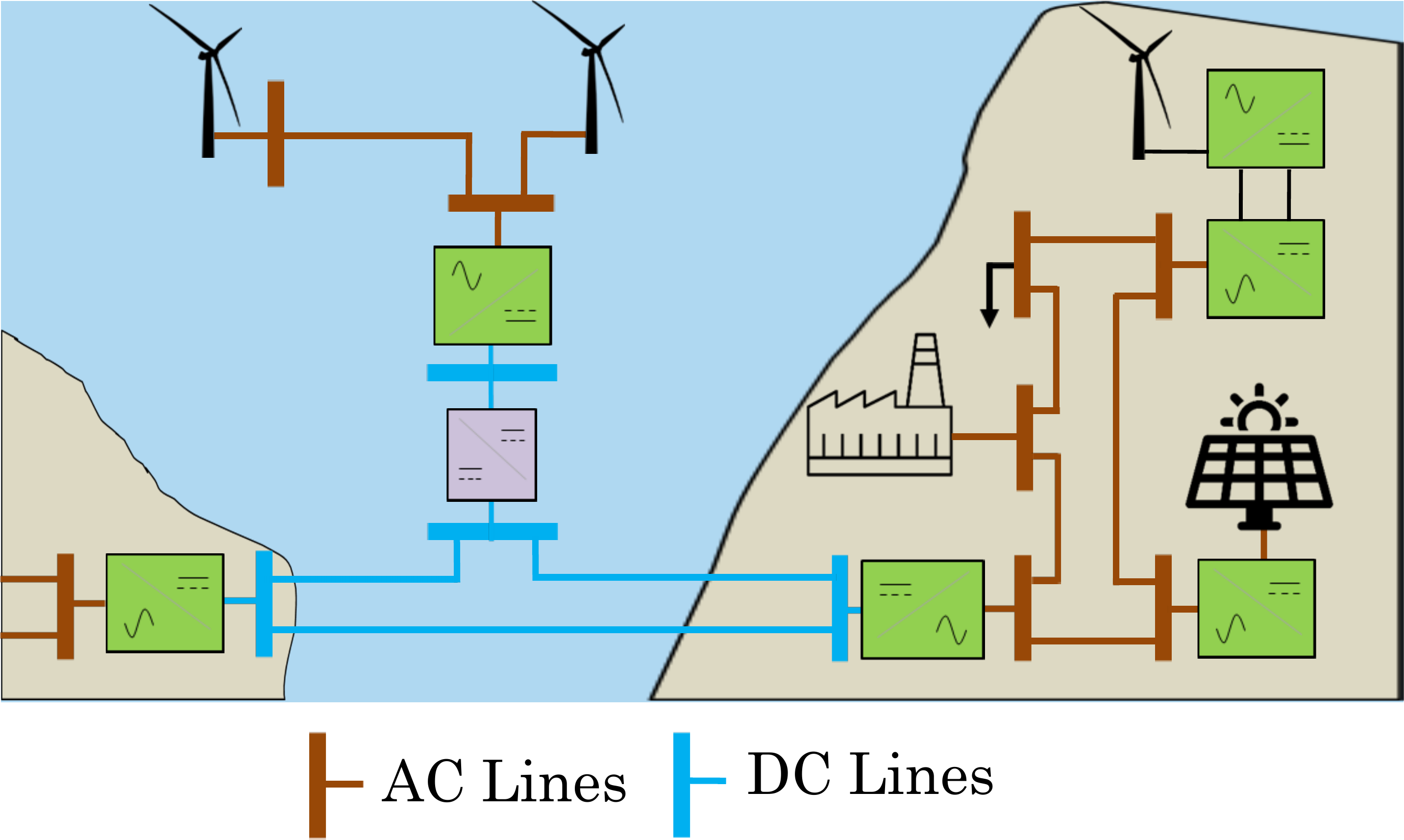}
   \caption{Example of converter-based power system.}
    \label{fig: pedps1}
\end{figure}

This article focuses on one of the possibilities for data-driven power system analysis: small-signal studies using data-based models and its possible applications. In particular, converter-based power system are considered (see Fig. \ref{fig: pedps1}). The ongoing progressive conventional synchronous generation (SG) displacement by converter interfaced generation still requires small-signal analysis in order to evaluate system stability
and the arised interactions \cite{hatziargyriou2020stability}. The high penetration of power electronic converters, due for example to the installation of large scale photovoltaic and wind power plants, affects system small-signal stability both in positive and negative manner, depending on several factors as the amount of SG power replaced, the network topology and the point of connection of the plants, among others \cite{hatziargyriou2020stability, shah2015review, 6513320, he2013small}. Both well known rotor angle stability and new stability phenomena, driven by converter controls interactions with other network equipment, need to be investigated through small-signal studies in order to ensure secure systems operation \cite{hatziargyriou2020stability}.

In literature, several studies concerning Machine Learning based small-signal stability assessment are provided. Most of them propose models for the damping ratio and the damped angular frequency prediction, as function of power injected by generators, power demand, lines' power flows, currents and buses voltages. Neural Networks (NN) and Deep Learning have been widely implemented \cite{feilat2007neural, teeuwsen2003small,cao2018deep,fu2020data}. In \cite{cao2018deep}, the analysis focuses on a power system characterized by high renewable generation penetration. In addition to the mentioned input features, solar and wind historical data are included, deploying the ability of a data-driven strategy to integrate information coming from different fields. In \cite{fu2020data}, a Convolution Neural Network is employed for the small-signal analysis of a large power system. It is proposed to use the obtained predictive model in an iterative manner, in order to compute preventive control actions. 
In \cite{teeuwsen2003small} and \cite{teeuwsen2005genetic}, the authors propose to use, respectively, NN and DT based algorithms for the estimation of the regions of the complex plane where the critical eigenvalues are located. DT was introduced in the field of power system analysis in \cite{wehenkel1989artificial} and \cite{wehenkel1994decision} for the assessment of system transient stability. It has been widely implemented in small-signal analysis as well. In \cite{zheng2012regression} two DT Regression based algorithms are implemented in order to estimate the voltage and oscillatory stability margins, respectively. The first is expressed as the distance between the maximum allowed power transfer and the actual power demand; the second is the damping ratio of the critical oscillation modes. In this study, data used for training are generated by PSS/E system simulations but the features selected can be easily accessible from real power system Phasor Measurement Units (PMUs). Thus, the DTs intrinsic ability to assess and show feature importance is deployed in order to suggest the optimal PMUs locations. In \cite{thams2017data} and \cite{venzke2020neural} the integration of data-driven stability assessment algorithms into Security Constraint Optimal Power Flow (SC-OPF) is proposed. In \cite{thams2017data}, DTs are trained to classify the system small-signal stability, using the damping ratio as metrics, for different fault contingencies and system operating points. Then, the decision rules estimated by the DTs, which express the power transfer limits that ensure small-signal stability, are added to the SC-OPF formulation.  In \cite{venzke2020neural} NN
%, through the use of non-linear relations (functions of the generator active power and voltage set-points), 
are used to encode the stability region in the SC-OPF, performing a stability classification based on the minimum system damping ratio. 

%The main contribution of the 
In this article a detailed methodology to develop a data-driven model to perform small-signal stability-related studies is described. 
While in \cite{feilat2007neural,cao2018deep,fu2020data,zheng2012regression,venzke2020neural} the use of data-driven techniques is limited to assess the small-signal stability, implementing a stable/unstable status classification or performing the regression of the critical modes damping ratios, in the present study a methodology for a complete small-signal analysis is provided. The objective of the algorithms developed is to predict the complete eigenvalue map and the Participation Factors (PFs), providing information about which network components have a dominant impact on the stability of the system. To the best of the author's knowledge, such methodology has not been covered in the literature.  Moreover, regression techniques have been exploited in order to deepen the knowledge of the studied systems. Conventional tools need a detailed system dynamics model and are able to provide information only about a particular operating condition per simulation. In contrast, regression-based models are trained in order to map the system response for a wide set of operating conditions.  Therefore, in this work it is proposed and described a possible way for organizing and visualizing the information drawn by the regression models, which allows improving the knowledge of the system behaviour.

%The article details 
The procedure to obtain the data driven model is detailed in the following sections, including: the requirements to develop a simulation model, the preparation and organization of the simulation model and corresponding mathematical equations to create the data-set, the data-driven model training phase and the final testing phase, to validate the results. This methodology is developed based on an essential test case and then tested on a slightly larger and more complex system in order to exemplify its application, but it could be expandable and applicable to larger models. The article also includes a discussion on the potential uses of the developed data-driven model.

The remaining parts of this paper are organized as follows. Section \ref{sec: methodology} describes the proposed methodology. Section \ref{sec: regression} provides a description of the regression techniques implemented. Section \ref{sec: testcase} illustrates the application of the proposed methodology to an essential power system test case. The performances obtained by the employed regression techniques are compared. Finally, a larger power system test case is studied. In Section \ref{sec: conclusion} the conclusions are drawn and some potential applications are identified.

\section{Methodology}
\label{sec: methodology}
\subsection{Problem Definition}
\label{subsec: conv_meth}
In general, power system dynamic response can be studied through time domain simulations. To do so, an accurate system modeling is required, typically based on non-linear differential and algebraic equations (DAE). Such equations are able to model the behavior of the network and the equipment installed in the system, including generators and converters, with their associated controllers, transformers, lines, cables and dynamic loads, among other devices \cite{kundur2007power}. %The set of equations resulting from power system modeling can be expressed in compact form as: % in \ref{ed: dae}.
The resulting set of equations can be expressed in compact form as: % in \ref{ed: dae}.
\begin{equation}
%    \alpha\dot{\textbf{x}}=f(\textbf{x}{,} \textbf{u}) \label{eq: dae}
    \dot{x}=f(x, u)
    \label{eq: diff}
\end{equation}
\begin{equation}
    y=g(x,u) \label{eq: alg}
\end{equation}

%\]
\noindent where equation (\ref{eq: diff}) represents the set of differential equations and equation (\ref{eq: alg}) the set of algebraic equations.
Vectors $x$, $u$, $y$ contain the state variables, the inputs and the outputs of the system. % Equations (\ref{eq: diff})-(\ref{eq: alg}) expresses the DAE which model network's equipment as generators and converters, with their controllers, dynamic loads and other devices. 
%The coefficient $\alpha$ assumes a value equal to zero when the equation is algebraic and equal to one when it is differential. Equation (\ref{eq: alg}) refers to the algebraic equations that model the network behavior.% The vectors \textbf{x} and \textbf{y} are the state and algebraic variables.% refer to buses voltage and to the remaining variables, respectively. $\textbf{V}_0$ and $\textbf{x}_0$ are their initial values, that can be obtained by power flows calculation of the static power system model or by state estimation. 
Conventional small-signal stability analysis involves the linearization of the system of DAE (\ref{eq: diff})-(\ref{eq: alg}), around an equilibrium point obtained via time-domain or power flow simulation. Through a first order Taylor's series expansion, the following generic linearized system, describing the temporal evolution of the inputs/perturbations, is obtained:
\begin{equation}
 \Delta \dot{x} = \textbf{A}\Delta x + \textbf{B}\Delta u %\tag{3}
 \label{eq: lin1} 
\end{equation}
\begin{equation}
     \Delta y = \textbf{C}\Delta x + \textbf{D}\Delta u %\tag{4} 
 \label{eq: lin2}
\end{equation}

Based on the previous model, the \textit{characteristic equation} of matrix \textbf{A} $ \in \mathbb{R}^{N\times N}$, with $N$ equal to the number of state variables, is obtained:
\begin{equation}
    \det (\lambda \textbf{I}-\textbf{A})=0
    \label{eq: cheqA}
\end{equation}

The roots of (\ref{eq: cheqA}),  $\lambda$ in $\mathbb{C}^M$, namely the eigenvalues of \textbf{A}, identify $M (=N)$ system's modes that are used to assess the stability of the system. The fact that the number of state variables is equal to the number of solutions of the characteristics equation ($N=M$) suggest the use of a common indexing. However, in the following, a distinct notation is maintained in order to indicate more explicitly the addressed variable. The eigenvalues can actually be both complex and real (here considered as complex values with null imaginary parts), indicating an oscillatory and a non-oscillatory response, respectively. 
Following Lyapunov's first method for small-signal stability assessment, the system results asymptotically stable if all the eigenvalues have negative real parts \cite{kundur2007power}. 

A deeper eigenvalues analysis, based on the computation of the PFs \cite{perez1982selective}, is used to extract information about the eigenvalues sensitivity. This analysis is based on the computation of the (square) participating matrix \textbf{P} $\in \mathbb{R}^{N\times M}$. Each element of \textbf{P}, namely the PF $p_{nm}$ is computed as:
\begin{equation}
    p_{nm}= \Phi_{nm}\Psi_{mn}
    \label{eq: pki}
\end{equation}

\noindent where $\Phi_{nm}$ is the $n$-th entry of the right eigenvector $\Phi_m$ (\ref{eq: righeigv}) and $\Psi_{mn}$ is the $n$-th entry of the left eigenvector $\Psi_m$ (\ref{eq: lefteigv}), where $n=1,...,N$ and $m=1,...,M$. 
\begin{equation}
    \textbf{A} \Phi_m = \lambda_m \Phi_m %\qquad %i=1,...,N
    \label{eq: righeigv}
\end{equation}
\begin{equation}
    \Psi_m \textbf{A}= \lambda_m \Psi_m % \qquad %i=1,...,N
    \label{eq: lefteigv}
\end{equation}

Each $p_{nm}$ measures the net participation of the $n$-th state variable on the $m$-th mode. Thus, from the analysis of \textbf{P} it is possible to estimate which state variables have a dominant impact on each mode and if interactions between different state variables occur.

The above described methodology for small-signal stability studies is widely used during power systems planning, design and operation phases. They are employed for checking the small-signal local and global stability for different power system operating points and load conditions and to verify that SGs and converters controllers are properly tuned, in order to avoid low damped oscillations \cite{kundur2007power}.

For large power systems, both the execution of time-domain simulations and the state-space linearization can be challenging tasks. Time-domain simulations are high-consuming in terms of both computation and time. Moreover, the detailed dynamics model required for the eigenvalue analysis can involve up to several thousands of state variables \cite{kundur2007power}, which complicates the information extraction from the model. 

Fast and low complexity models can be obtained by the use of data-driven and data-mining techniques, as proposed in the following sections. In addiction, such data-based models can further unlock additional uses and applications, such as: additional information extraction, inter-variable relation, IP-protected model development, stability indicators calculation, data-driven optimization including stability constraints, among others. Further discussions about these applications can be found in Section \ref{sec: conclusion}.

\subsection{Proposed Data-Driven Methodology}
\label{subsec: new_meth}

\begin{figure}[!ht]
\centering
    \includegraphics[width=\linewidth]{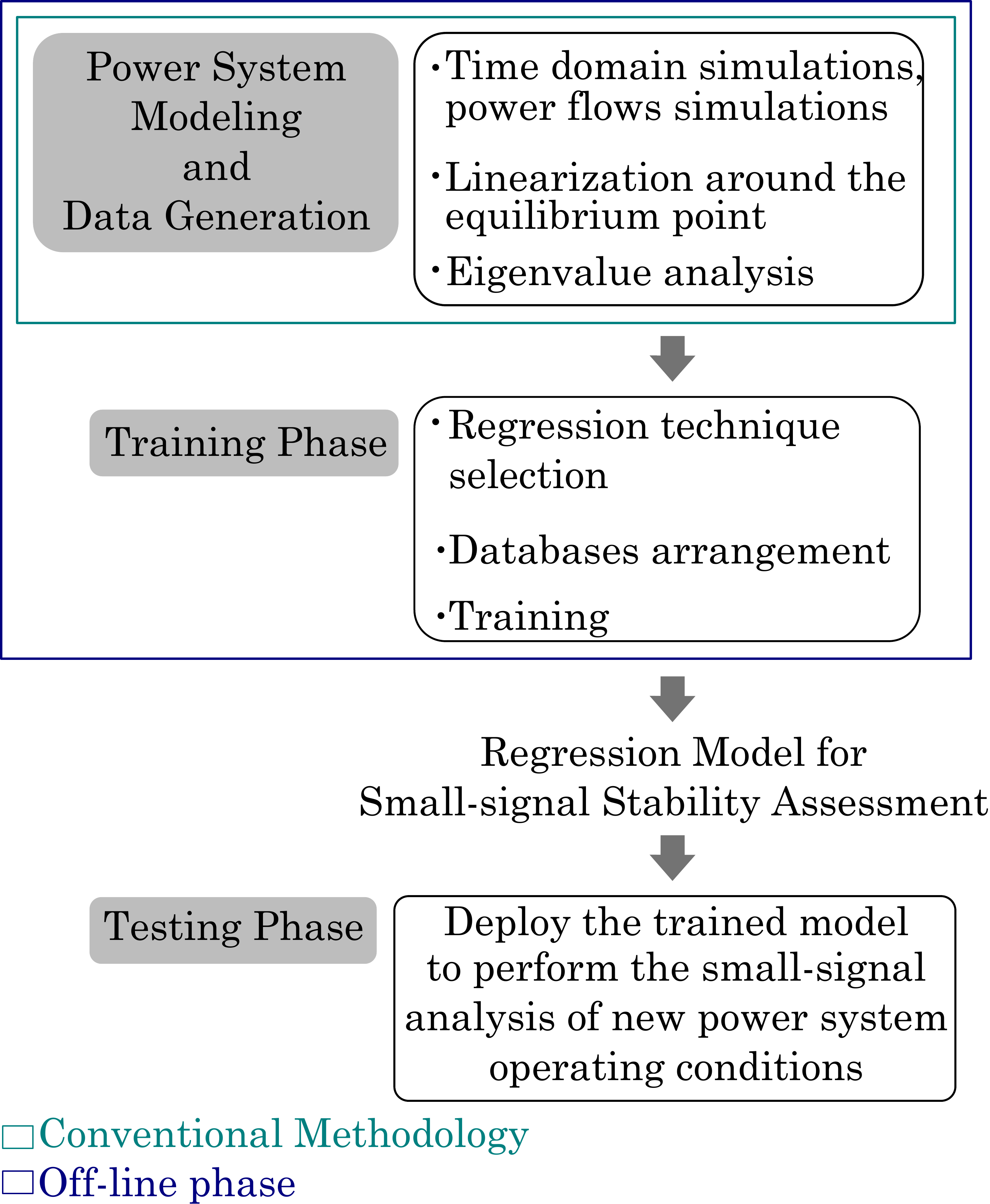}
    \caption{Flow-chart of the proposed methodology.}
    \label{fig: flowchart}
\end{figure}

The study presented in this paper proposes a methodology for power system small-signal stability analyses, using regression techniques, as complementary to time-domain simulations and eigenvalue analysis (see Fig. \ref{fig: flowchart}). The procedure defined by the proposed methodology involves the following steps to obtain a data-driven system model:
\begin{itemize}
\item \textit{Power system modeling}: build the power system model in the selected simulation package environment (e.g. PSS/E, DigSilent, PSCAD, EMTP-RV, Matlab Simulink, etc.).
\item \textit{Data generation}: for several values of the parameters of interest and for different system configurations, obtain operation points from time-domain simulations (or power flow calculations) and compute eigenvalues and PFs from the state-space model.%execute time-domain simulations and state space-linearization for several values of the parameters of interest and collect the results (the inputs and outputs of the data-driven model, respectively).
\item \textit{Regression technique selection}: choose the regression technique to implement. The choice is driven by several factors: the linear o non-linear nature of data to be fitted, the desired level of accuracy, robustness, scalability, interpretability and complexity of the algorithm. 
\item \textit{DBs arrangement}: organize the obtained results in training DBs with a pairwise input-output structure. It has to be suitable and consistent with the regression strategy to be implemented.
\item \textit{Training}: make the regression model able to approximate the selected outputs, as function of the inputs. Tune the parameters of the model in order to achieve good performances, avoiding overfitting \cite{friedman2001elements}.
\item \textit{Testing}: deploy the obtained regression models for predicting the output values related to new input instances, not belonging to the training DB. The performance of the model is evaluated in terms of accuracy and computing time, comparing the prediction with the corresponding results obtained by the conventional methodology.
\end{itemize}

The \textit{Power system modeling} and \textit{Data generation} steps correspond to the implementation of the conventional methodology described in Section \ref{subsec: conv_meth}. The suggested data-driven methodology is based on the fact that small-signal studies are typically carried out following approaches analogous to the one described in the prior section. It also considers that power system models, as well as simulation models and results, are already available for engineers who work in power system stability assessment. %Besides, the execution of the simulations, which is the most computationally intensive task, is limited to the "off-line" phase, which involves also the regression model training, and precedes the use of the model for assessing the stability of new instances. Thus, the computational intensive phase and the fast model deployment are temporally separated.  

The \textit{Training} and \textit{Testing} phases are the core of the data-driven part of the procedure. It is worth to consider two aspects that makes the use of regression model advantageous: The time decoupling between  the model preparation and model deployment phase and the additional freedom in setting up the model.  In fact, the steps carried out to build the data-driven model, i.e. the simulations execution and the regression training, which represents the most computationally expensive tasks, are conducted in a phase that can be considered as a preparatory step, off-line with respect to the phase in which the model is used. The trained model, which contains information about the system response to different operating conditions, is then asked to predict the system status for unknown instances during testing. In this phase, the low-demanding computing properties of regression-based model are exploited. 

Concerning the modeling setup, regression-based model differs from conventional methodology on the number of independent variable needed for realizing an accurate modeling. A small-signal analysis involves the formulation of several equations and considers an extensive number of variables in order to properly capture the dynamics of the system equipment. On the contrary, regression-based approach uses a limited number of selected independent inputs variables, interchangeably referred to as features. The advantage does not stay only in the reduction of the number inputs, but also in introducing the possibility of selecting them in an user's need oriented manner. The features selection is extended also to those variables which are not linked to the outputs by clear mathematical equations. In fact, features can be also chosen on the basis of their availability and accessibility, as measured quantities, or because of particular interest to the user. Obviously, the degree of freedom for the selection is limited by the accuracy achieved by the predictive model, which can be considered as a trade-off objective and as an indicator of the correlation quality between the selected inputs and the outputs.

\section{Regression Techniques}
\label{sec: regression}
Regression problems can be solved employing a wide variety of  techniques. A first important selection is the nature of the model to be used: linear or non-linear. The selection of the regression technique to implement is heavily influenced by the objective to be achieved. For this reason, the accuracy and robustness of the model are not the only characteristics to be considered. Depending on the study objective and the application field, features such as: computing time, scalability and interpretability might become especially relevant. Also, as discussed in \cite{kamwa2011accuracy}, data-mining models are affected by a trade-off between accuracy and transparency. The need of transparent, interpretable and explainable machine learning algorithms is emerging since they allow the user verifying and interpreting how the algorithm works \cite{carvalho2019machine}.

In this work the small-signal analysis, i.e. the evaluation of the modes and of the PFs of a power system subjected to a small perturbance, is dealt as a non-linear regression problem. In the choice of the regression technique, characteristics as transparency and interpretability have been preferred, both concerning the model development, the formulation of the models and the presentation of the results. Therefore, the techniques implemented are DT and Spline-based regressions. 

DT-based regressions are simple to develop, easily interpretable and allow to implement a multi-output strategy. Spline-based regressions are based on non-linear interpolations and approximations through piece-wise polynomial functions. They allow to manage non-linear problems with models that include low complex functions (i.e. with low polynomial degree). In the following subsections a description of their basic principles is provided, as well as concepts and definitions useful to understand how these techniques have been implemented in the test case, as described in Section \ref{sec: training}.

\subsection{DT Regression}
\label{sec: dt}
DT Regression is a decision algorithm with a tree-like structure able to predict the value of a continuous numerical target through a series of recursive DB splits \cite{friedman2001elements}.
In this study, DT regressions have been developed using the Python library Scikit-Learn \cite{scikit-learn}, which implements a version of the Classification And Regression Trees (CART) algorithm \cite{breiman1984friedman}. 
\begin{figure}[!ht]
    \centering
    \includegraphics[width=\linewidth]{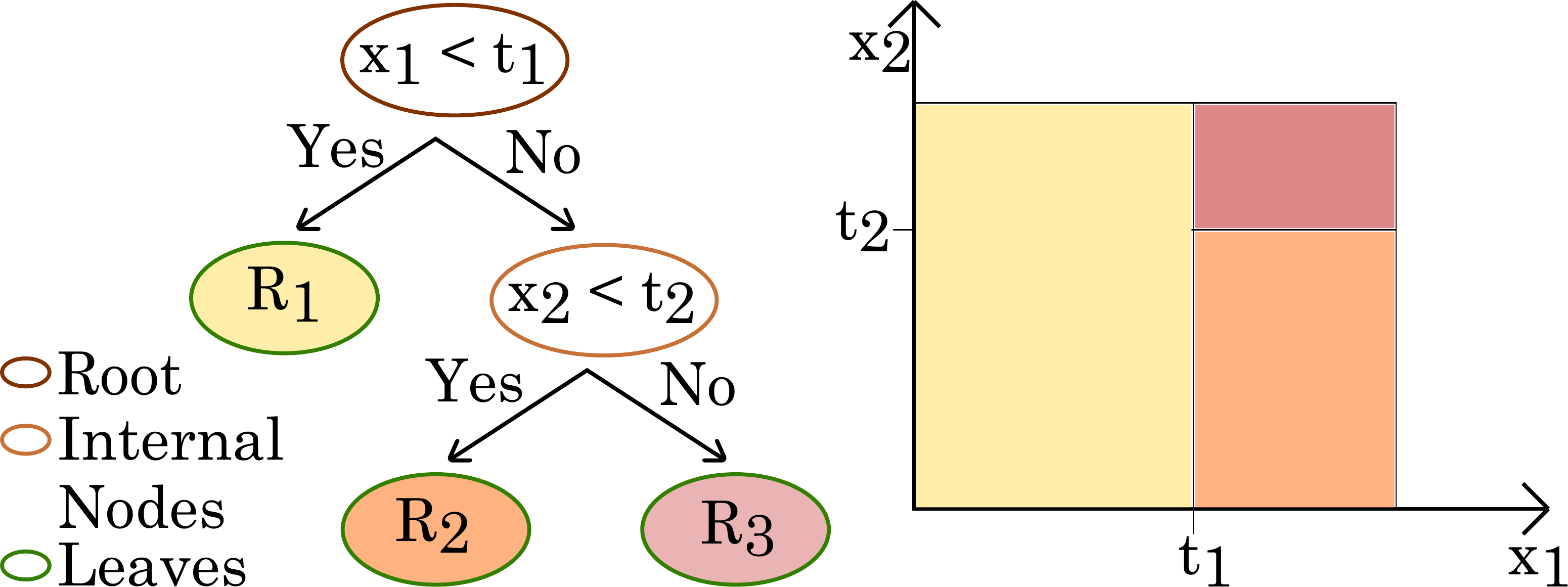}
    \caption{Example of the structure of a simplified trained DT regression (on the left) and of the related feature space partition (on the right).}
    \label{fig: tree_example}
\end{figure}

The left side of Fig. \ref{fig: tree_example} shows an essential scheme of the structure of a DT. The first node of the tree-like structure is called \textit{root}: it contains all the samples of the DB and generates the first couple of child nodes. Weather the child nodes undergo further splits, they are called \textit{internal nodes} or \textit{leaves}. Thus, the \textit{leaves} are the terminal nodes which provide the regression outcomes ($R_1$, $R_2$ and $R_3$, in Fig. \ref{fig: tree_example}). The outcome can be a single value or a set of values, weather a single-output (SO) or a multi-outputs (MO) approach is used.

At each node split, the samples contained in the node are divided into two subsets. They are associated to the proper child node depending on the adopted splitting rule. The splitting rules are expressed as operational thresholds, applied to the features of the problem, as $x_1<t_1$ in Fig.~1a, where $x_1$ is a feature and $t_1$ is a numerical threshold value. The choice of the feature involved by the splitting rule and the computation of the threshold value depend on the selected splitting criterion and on how it is applied. The splitting criterion is based on the minimization of the weighted average of the \textit{impurity} of the child nodes’ samples. Usually, in regression problems, the \textit{impurity} is measured as the Mean Squared Error (MSE) or as the Mean Absolute Error (MAE). Besides, the splitting criterion can be applied in order to realize the \textit{best} split or the \textit{best random} split. In case the \textit{best} split is desired, among all the candidate splits $\theta=(x,t)$, the splitting rule is determined by $\theta$ which returns the minimum \textit{impurity}. Otherwise, the \textit{best random} split randomly selects the feature to which apply the operational threshold and estimates the value of $t$ in order to minimize the \textit{impurity}. 

Moreover, it is possible to use a single tree or an ensemble model. There are several strategies for the implementation of ensemble DT regression. In this study the bagging strategy has been adopted \cite{friedman2001elements}, \cite{ernst2005tree}, which consists of training, for the same regression problems, several different DTs. Then, the result of the regression is computed as the average of the outcomes of all the trained DTs. This strategy is able to improve the accuracy reducing the variance of the model \cite{friedman2001elements}.

Concerning the scalability, a fully-grown CART algorithm has a computational complexity tending to $O(N_f N_s \mbox{log}N_s$), where $N_f$ is the number of features and $N_s$ the number of samples of the training DB \cite{aluja2003stability}. Usually, the model complexity is limited in order to reduce the training runtime as well as to avoid overfitting and to improve the estimator generalization ability. The hyperparameters tuning process, through a grid-search cross validation strategy, and the \textit{minimal cost-complexity pruning} are applied with this aim, as described in \cite{friedman2001elements}. 
Once the DT structure is built, i.e. the DT regression is trained, it is possible to compute the regression output for features not belonging to the training DB. Depending on the values of the features, a unique path, from the root to one of the leaves, is individuated. The regression output is calculated as the average value of the samples contained in the leaf at the end of the path.

In addiction to the prediction of outcomes for unknown instances, DT regression models provide further information useful to deepen the carried out analysis. An example can be explained considering Fig. 3. The tree-like structure on the left and the feature space partition on the right refer to the same regression problem but point out different aspects. The tree structure highlights which splits have been computed and return information about feature importance. The feature importance is related to the predictive ability of the feature and, since the features of the upper levels of the tree are involved in the splits of the larger fractions of samples, they result in higher importance. It is worth to consider that this kind of feature importance estimation depends on the training DB structure and tends to favor that features that have higher cardinality, i.e. the ones that are present with a larger number of unique values. However, with the proper, balanced DB composition it can extract useful information, as in \cite{zheng2012regression}. The feature space partition is suitable for mapping the value of the output variable, pointing out which range and combination of features values bring to a desirable or undesirable output value.

\subsection{Spline Regression}

Splines are used to interpolate or approximate a non-linear function through piece-wise polynomial curves. In this work spline regressions have been implemented using the Matlab Spline Toolbox \cite{mat_spl_toolbox}. For a detailed mathematical background refer to \cite{de2005spline} and \cite{de1978practical}.

\begin{figure}[!ht]
    \centering
    \includegraphics[width=\linewidth]{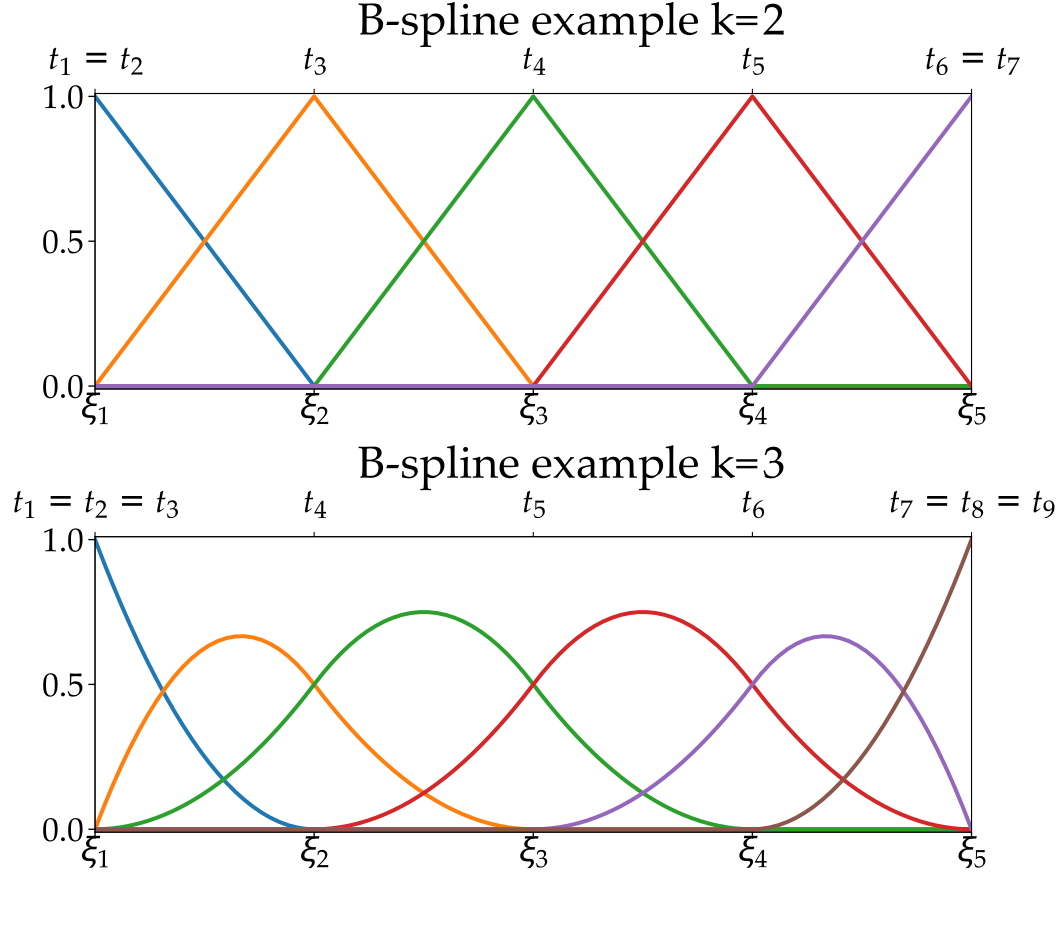}
    \caption{Example of 2nd order and 3rd order B-splines.}
    \label{fig: bspline}
\end{figure}

For the spline construction it is common to use the B-form, which fits the curves as a weighted sum of basis splines (B-splines), over the interpolation interval. An univariate spline in the B-form is expressed as
\begin{equation}
    f(x)= \sum_{j=1}^{J} B_{j,k}(x) a_j %\qquad i=1,...,n
    \label{eq: univ_sp}
\end{equation}
\noindent where $J$ is the number of adjacent segments which compose the interpolation interval, $B_{j,k}$ is the $j$-th B-spline of order $k$ and $a_{j}$ is the coefficient to be fitted in order to realize the linear combination of $J$ B-splines that interpolate or approximate the known data points. B-splines of 2nd and 3rd order are shown in Fig. \ref{fig: bspline}, as an example. %Fig.  shows B-Splines of 2nd and 3rd order, which correspond to $B_{j,2}(x)$ and $B_{j,3}(x)$, respectively. 
Each $B_{j,k}$ is defined over a knot sequence $t:=(t_a)_{1}^{J+k}$, it is non-negative within the interval [$t_j$,...,$t_{j+k}$] and equal to zero outside it. A knot sequence is an increasing sequence of points, obtained \textit{augmenting} the break points sequence ($\xi:=(\xi_b)_{1}^{J+1}$), made by the points that define the segments of the interpolation interval. In particular, the boundary break points are included $k$ times in the knots sequence, while the internal break points  can be repeated, modifying the continuity and the smoothness of the fitted curve in that point. For instance, the B-splines of Fig. \ref{fig: bspline} have knot sequences (shown by the upper x-axes) with internal knots multiplicity equal to one (simple knots): this makes $B_{j,2}(x)$ continuous and $B_{j,3}(x)$ continuous, with a continuous first order derivative at the break points $\xi$.  

Univariate splines can be built both with interpolation and approximation approaches, fitting a curve that passes through the known points or that minimizes the least square difference with the known points, respectively.
In case of univariate spline approximations, it is possible to estimate the optimal knots position, with the use of the \textit{newknt} function provided by the Matlab Spline Toolbox \cite{mat_spl_toolbox}. Hence, at the beginning an univariate spline approximation is computed using a sequence of uniformed distributed knots. Then, the knots sequence is recalculated in order to optimize the approximation. 

Computing the tensor product of univariate splines, multivariate splines can be obtained. Bivariate splines are expressed as 
\begin{equation}
    f(x,y)= \sum_{j=1}^{J}\sum_{v=1}^{V} B_{j,k}(x) B_{v,l}(y) a_{j,v}
    \label{eq: biv_sp}
\end{equation}
\noindent where $k$ and $l$ are the orders of the spline along the $x$ and $y$ directions, respectively. 

In order to obtain the formulation in (\ref{eq: biv_sp}), the two-step procedure illustrated in \cite{de2005spline}, aimed to fit the coefficient $a_{u,v}$, can be carried out. Let consider $z \in \mathbb{R}^{Nx \times Ny}$ known points. The objective is to fit the $f=(x,y)$ function that approximates the  $z$ points. The first step is aimed to the estimation of one spline (of order $k$) for each of the $N_y$ data sets as function of $x$.
The group of univariate splines as function of $x$ has the same formulation in (\ref{eq: univ_sp}). The second step consists in the approximation of the values $a_j$ through a set of univariate splines as well. This time the approximation is conducted as function of $y$ and results in a set of univariate splines of order $l$:
\begin{equation}
    a_j= \sum_{v=1}^{V} a_{j,v} B_{v,l}(y)
\end{equation}

This procedure can be applied also inverting the directions along which the splines are calculated in the two steps and leads, in both cases, to the formulation in (\ref{eq: biv_sp}).

\section{Test Case}
\label{sec: testcase}

In this section the application of the proposed data-driven small-signal model is presented  for two power system examples. The procedure delineated in Section \ref{subsec: new_meth} is first applied to a 3-bus power system, which main characteristics are provided in Section \ref{subsec: testcase1}, but for a more detailed description refer to \cite{collados2019stability}. Sections from \ref{subsec: datagen} to \ref{subsec: testingph} illustrate the proposed methodology's steps, dealing with the application of both the regression technique described in Section \ref{sec: regression}. Their performances are compared and the one that achieves the best results has been applied to a still essential 9-bus power system, with an higher complexity due to the presence of two SGs and eight loads (see Section \ref{subsec: testcase2}).

\subsection{Power System Description}
\label{subsec: testcase1}
\begin{figure}[!ht]
    \centering
   \includegraphics[width=\linewidth]{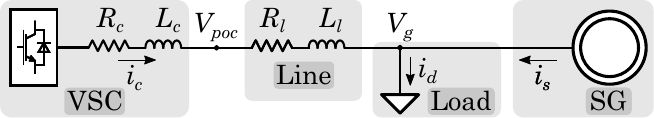}
    \caption{Scheme of the simplified model with aggregated thermal-based generation (SG), aggregated VSC-based components (VSC) and aggregated power demand (Load) \cite{collados2019stability}.}
    \label{fig:PSmodel}
\end{figure}

The Power System object of the small-signal stability analysis has been previously modeled and studied in \cite{collados2019stability}. The system represents a simplified and aggregated model of an AC grid with high penetration of Voltage Source Converters (VSCs). As shown in the scheme of Fig. \ref{fig:PSmodel}, a single converter is used to model the VSC-based elements (possibly aggregating HVDC links, wind and photovoltaic power plants), the thermal-based generation is represented by an equivalent SG and the aggregated power demand is represented by a load.

The system is modeled by a complete non-linear model, developed in Matlab Simulink. The dynamics of the SG and of the VSC are modeled in a detailed manner including: the SG model is composed of a generator model, an excitation system, a steam turbine and a governor, including a frequency droop control; the VSC model is composed of a PLL control, an inner current control loop, a power control, a voltage control and a frequency droop control. All the details about the system modeling are provided in \cite{collados2019stability}.

\begin{figure}[!ht]
    \centering
    \includegraphics[width=\linewidth]{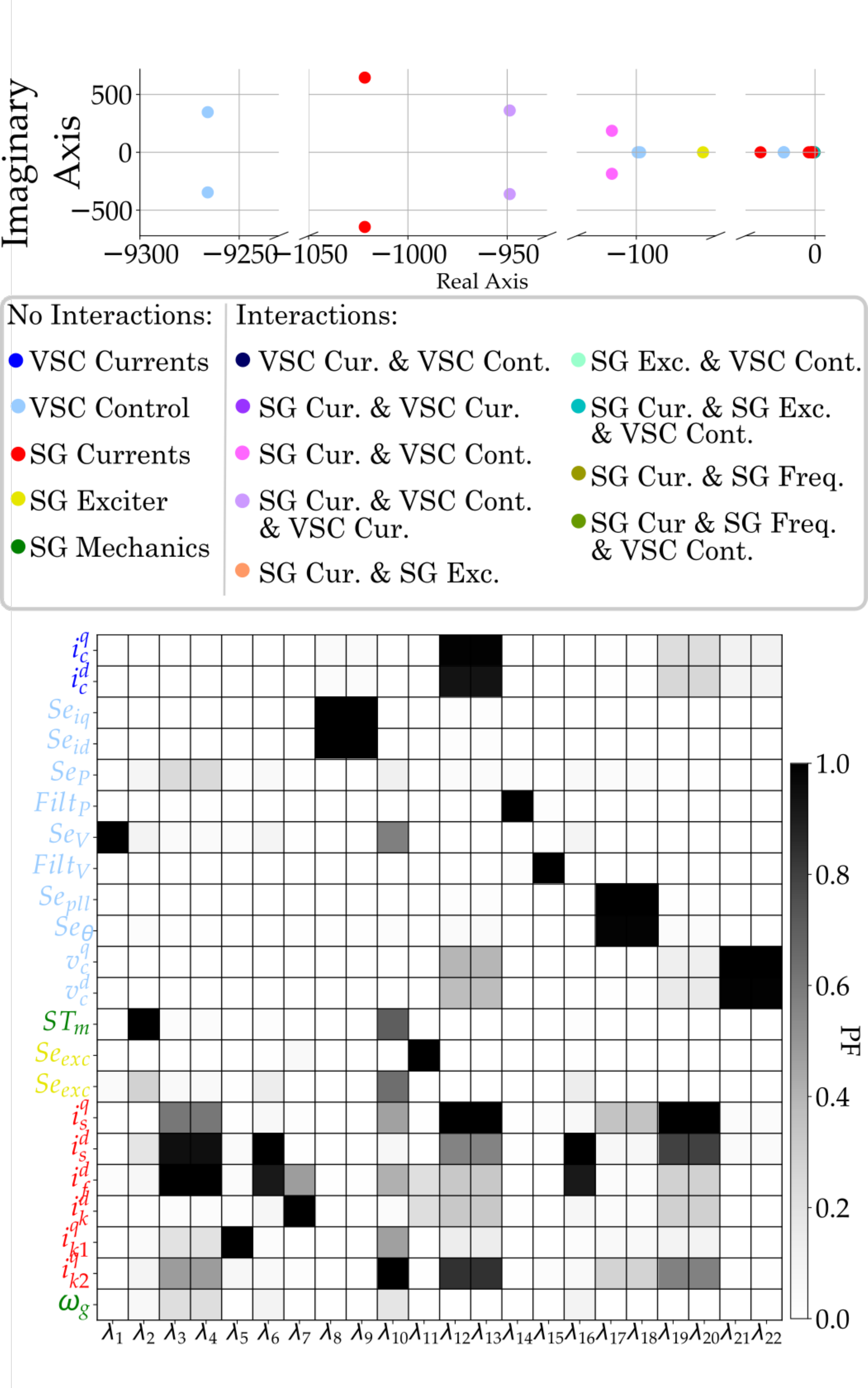}
    \caption{By way of example, the results of the small-signal analysis related to a single simulated case. Top panel: modal map. Second panel: legend of colors used to indicate the  poles' dominant $P_{var}$ group. Bottom panel: PFs matrix; the variables on the vertical axis are the problem state variables. }
    \label{fig: modalmap}
\end{figure}

In \cite{collados2019stability}, a small-signal stability analysis of the system is conducted in order to identify the impact of the VSC controllers and the minimum SG rated power required to ensure stability. The procedure for the estimation of the modal map and of the PFs follows the conventional methodology described in Section \ref{subsec: conv_meth}. The equilibrium points used in the state-space linearization are obtained by time domain simulations, executed for several values of: VSC controllers’ parameters and SG rated power. A small-signal perturbation is considered as an input for the system. An example of modal map and of PFs matrix related to a single simulated condition is shown in Fig. \ref{fig: modalmap}. The state-space model and the eigenvalue analysis identify $M$=22 system oscillation modes ($\lambda$), related to $N$=22 state variables. The state variables, namely the participating variables ($PVs$) in the PFs analysis, are related to the VSC and SG controllers: they are grouped in variables related to VSC currents, VSC controllers, SG mechanics, SG Exciter and SG currents. For each pole, the PFs of each PV are computed. The PFs are normalized values and the value of 0.3 is chosen as the lower threshold that indicates if the PV has a relevant impact on the mode. From the PFs analysis emerges that a pole can be subjected to the dominant influence of a single PV or of a combination of them, revealing interactions when PVs of the VSC and SG controllers are simultaneously dominant for a pole. For an easy graphical representation of the PFs analysis’ results on the modal maps, a color is associated to each pole that highlights the PV with higher participation (Fig. \ref{fig: modalmap}).

\subsection{Data generation}
\label{subsec: datagen}

A number of parameters are involved in the dynamics of the equipment and affect the system operating conditions. In the scope of this work, in order to exemplify the proposed methodology, three parameters have been selected: the SG rated power, which is a physic variable, and the VSC voltage control and the VSC frequency droop control, which are control variables. In particular, the latter two are the voltage controller time constant $\tau_v$ and the frequency droop characteristic $R=\Delta \omega / \Delta P$, where $\Delta \omega$ is the frequency variation and $\Delta P$ is the active power variation. Therefore, simulations have been conducted for different values of the SG rated power, keeping constant the VSC rated power (equal to 500 MVA). The load demand is set equal to 500 MW at the beginning of the simulation and, at time equal to 5 s, a load variation of 1\% is imposed in order to simulate a small disturbance. The simulations have been conducted setting the following values of SG rated power, $R$ and $\tau_v$:
\begin{itemize}
    \item SG rated power has been set equal to 500 MVA, 400 MVA, 300 MVA and 200 MVA;
    \item $\tau_v$ has been set equal to 0.01 s, 0.05 s, 0.1 s, 0.3 s, 0.5 s, 0.7 s and 1 s;
    \item \textit{R} has been set between 0.01 and 1, with a step of 0.01, has been executed.
\end{itemize}
Therefore, the total amount of simulations is $N_s= 4 \times 7 \times 100= 2800.$ For each simulation instance, the poles real and imaginary parts and the related PFs are extracted.

\subsection{DBs Arrangement}
\label{sec: dborg}
Two regression problems need to be solved: the first deals with the prediction of the poles location; the second deals with the prediction of the PFs. The poles' regression problem is split in two sub-problems, one related to the computing of the real part of the poles and one related to the imaginary part. Thus, the results of the simulations are organized in three main DBs: one for the regression of the poles imaginary part ($\textbf{DB}_{\Im}$), one for the regression of the poles real part ($\textbf{DB}_{\Re}$) and one for the regression of the PFs ($\textbf{DB}_{PF}$). Each DB has a pairwise inputs-outputs structure as shown in (\ref{eq: dbpoles}) and (\ref{eq: dbpf}).
\begin{equation}
\textbf{DB}_{\Re/\Im}\in \mathbb{R}^{N_s\times(N_f+M)} = 
\left( \begin{array}{@{}c|c@{}}
    \textbf{X} & \textbf{Y}_{\Re/\Im}\\
    \cmidrule[0.4pt]{1-2}
   \begin{matrix}
   X_{1} \\
    \hdotsfor[2]{1}\\
    X_{i} \\
    \hdotsfor[2]{1}\\
    X_{Ns}
    \end{matrix}
    & \begin{matrix} Y_{\Re/\Im,1} \\
    \hdotsfor[2]{1}\\
    Y_{\Re/\Im,i} \\
    \hdotsfor[2]{1}\\
    Y_{\Re/\Im,Ns}
    \end{matrix} \\
   %\cmidrule[0.4pt]{1-2}
\end{array} \right)
\label{eq: dbpoles}
\end{equation}
%The complete DB for the PFs regression problem, $\textbf{DB}_{PF}$, is:
\begin{equation}
    \textbf{DB}_{PF}\in \mathbb{R}^{(N_s \cdot N)\times(N_f+M)} = 
\left( \begin{array}{@{}c|c@{}}
\textbf{X} & \textbf{Y}_{PF}\\
    \cmidrule[0.4pt]{1-2}
   \begin{matrix} X_{1} \\
\hdotsfor[2]{1}\\
 X_{i} \\
 \hdotsfor[2]{1}\\
 X_{Ns}
\end{matrix}
      & \begin{matrix} \textbf{Y}_{PF,1} \\
\hdotsfor[2]{1}\\
 \textbf{Y}_{PF,i} \\
 \hdotsfor[2]{1}\\
 \textbf{Y}_{PF,Ns}
\end{matrix} \\
   %\cmidrule[0.4pt]{1}
\end{array} \right)
\label{eq: dbpf}
\end{equation}

The inputs matrix $\textbf{X}$ in (\ref{eq: dbpoles}) and (\ref{eq: dbpf}) collects the value of the parameters $S_{SG share}$, $\tau_{v}$ and $R$ involved in the executed time-domain simulations. Therefore, $\textbf{X} \in \mathbb{R}^{N_s\times N_f}$, with $N_f=3$. Concerning the SG rated power, it is expressed as the share of SG installed in the system, evaluated as
$S_{SG share}= S_{SG}/(S_{SG}+S_{VSC})$, where $S_{SG}$ and $S_{VSC}$ are the SG and VSC apparent rated power in the network, respectively. In this way data standardization can be avoided.

The matrices of the outputs of the real part and imaginary part regression problems, $\textbf{Y}_{\Re}$ and $\textbf{Y}_{\Im}$, collect the real and imaginary parts of all the poles for each simulation instance.  The matrices are $\textbf{Y}_{\Re}$, $\textbf{Y}_{\Im}$ $\in$ $\mathbb{R}^{N_s\times M}$.

Concerning the PFs regression problem, for each $i$-th simulation instance a PFs square matrix $\textbf{Y}_{PF,i}$ $\in \mathbb{R}^{N\times M}$ is generated. Hence, the complete DB contains $N_s$ PFs matrices and the resulting complete output matrix is $\textbf{Y}_{PF} \in \mathbb{R}^{N\times M \times N_s}$ (or $\textbf{Y}_{PF} \in \mathbb{R}^{(N_s\cdot N) \times M}$ if seen from a bidimensional point of view).

\begin{figure}[!ht]
    \centering
    \includegraphics[width=\linewidth]{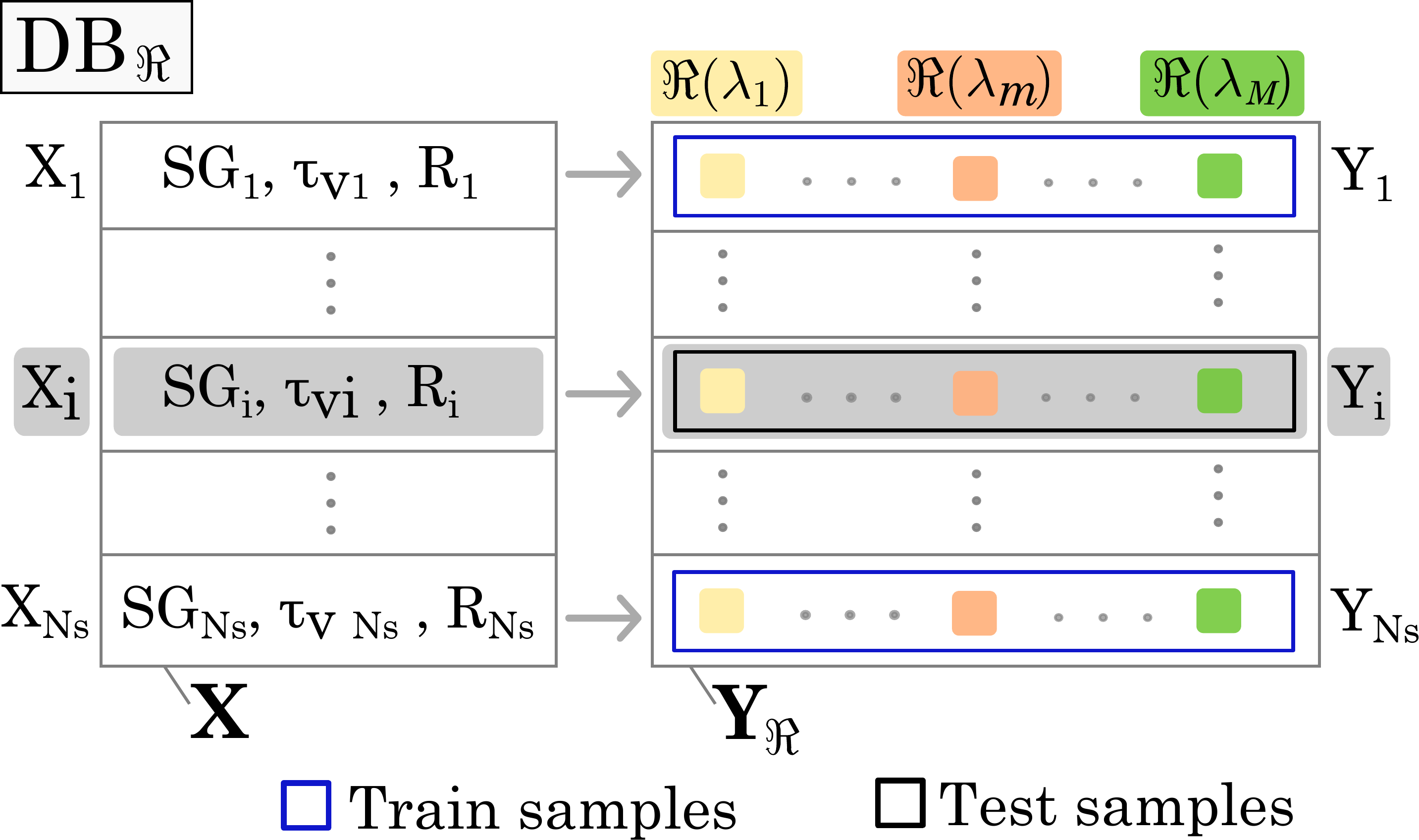}
    \caption{DB for the MO regression of the poles' real part.
    }
    \label{fig: db_remo}
\end{figure}

\begin{figure}[!ht]
    \centering
    \includegraphics[width=\linewidth]{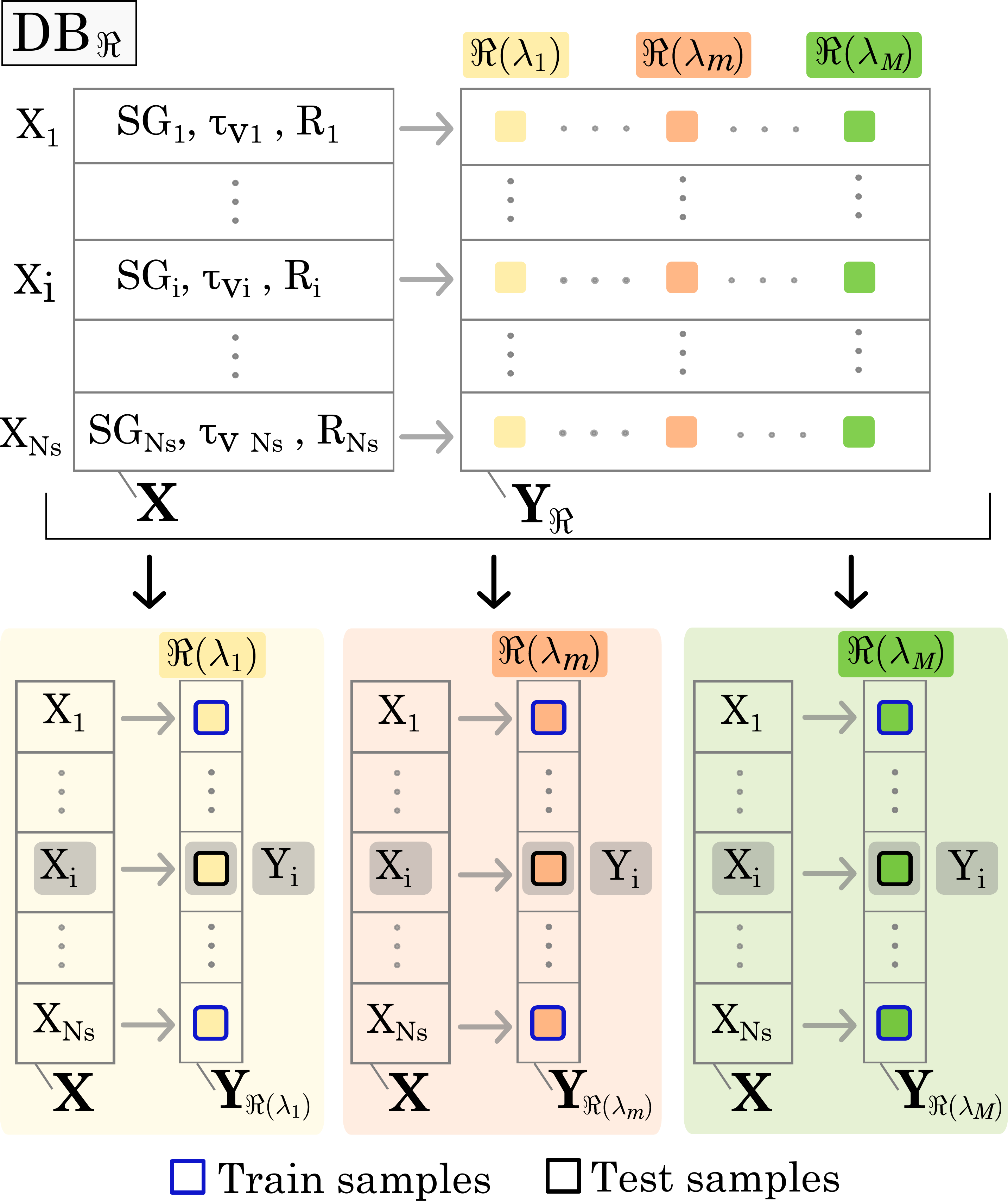}
    \caption{DBs for the SO poles real parts regression.}
    \label{fig:db_reso}
\end{figure}

\begin{figure}[!ht]
    \centering
    \subfloat[]{\includegraphics[width=\linewidth]{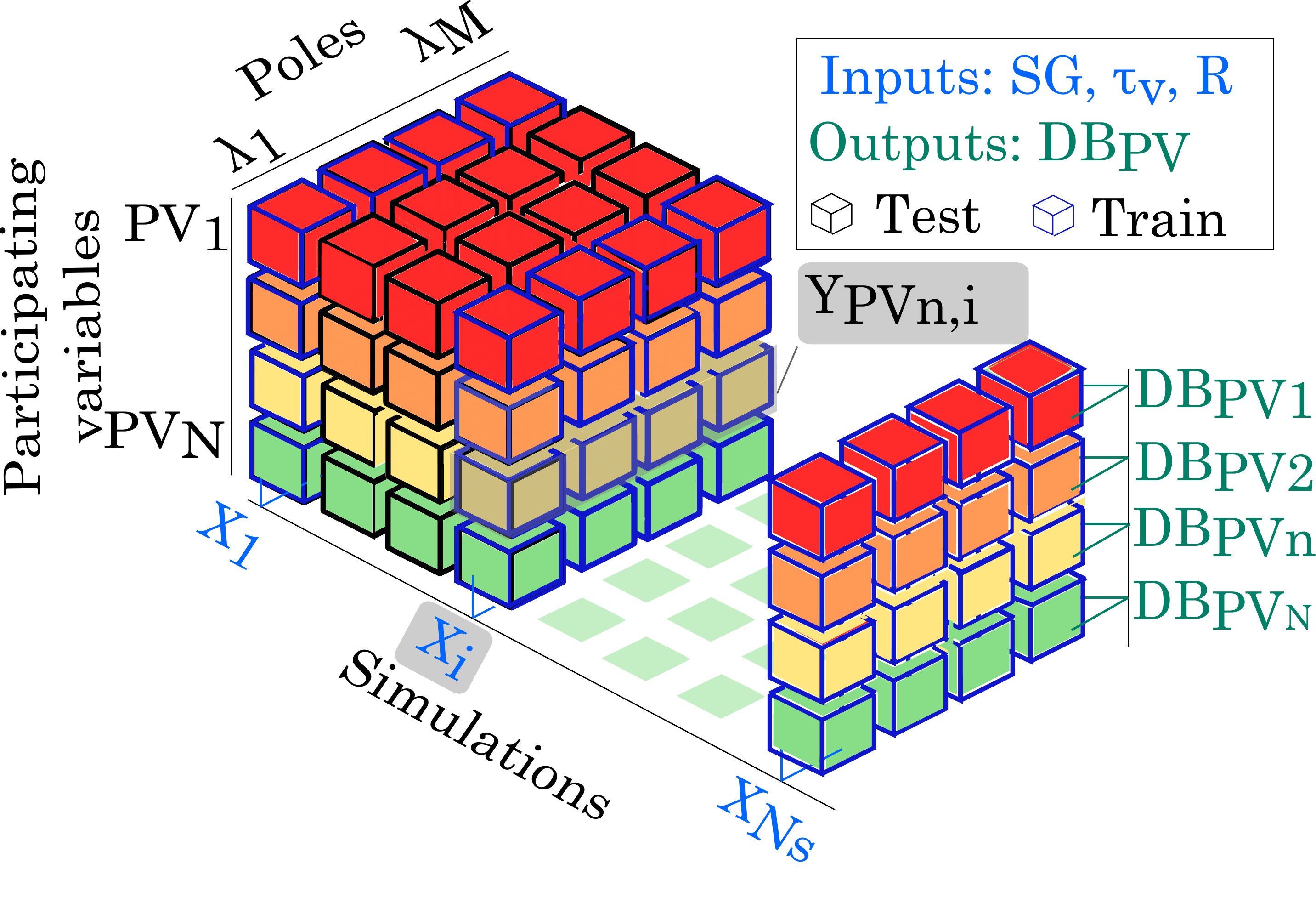}
    \label{fig: db_pfvar}
    }
    \hfil
    \subfloat[]{
    \includegraphics[width=\linewidth]{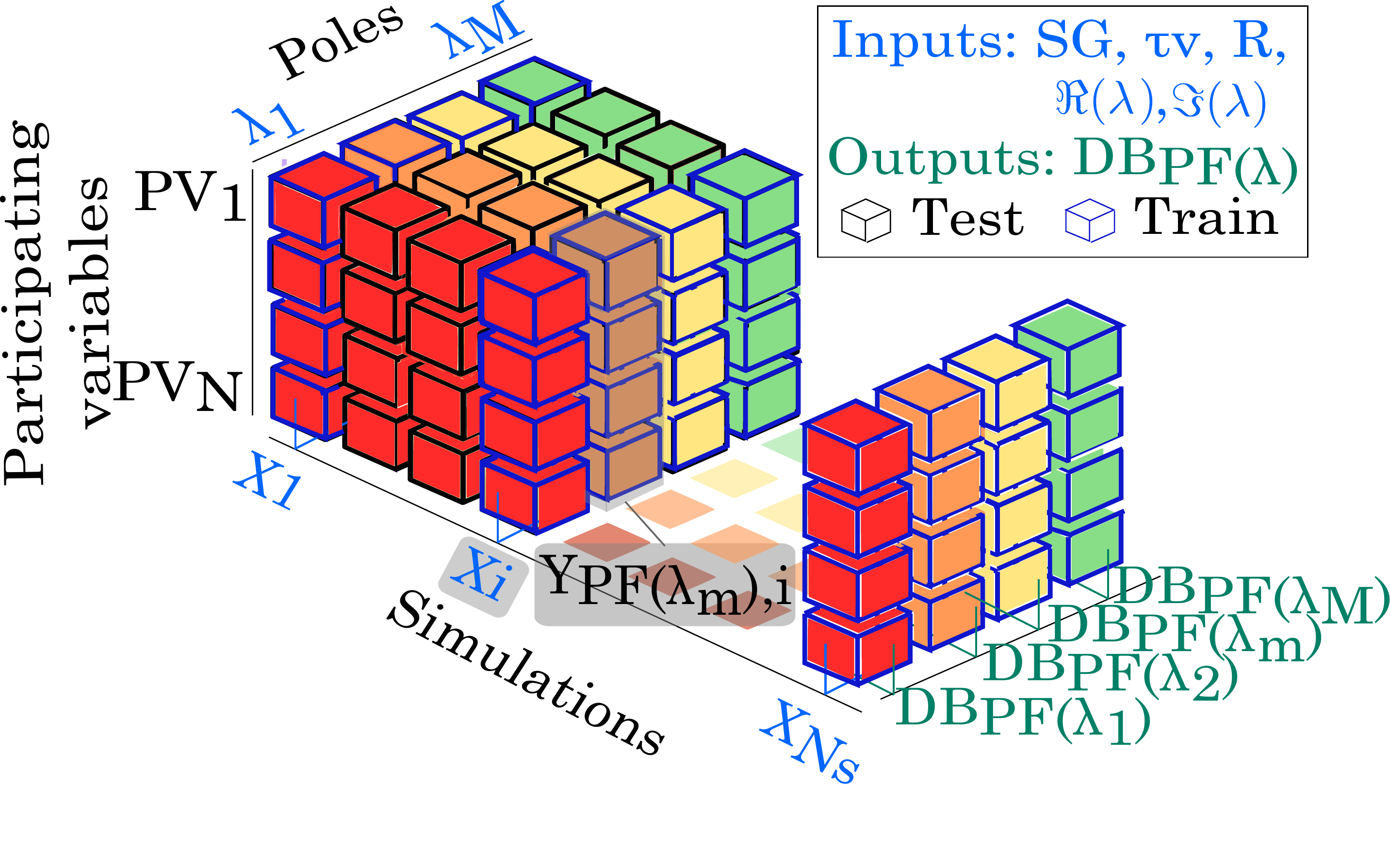}
    \label{fig: db_pfpole}}
    \caption{The whole 3-D structures ($N\times M\times N_s$) represent $DB_{PF}$. Fig. \ref{fig: db_pfvar} shows the $DB_{Pvar}$ subsets, used for the PFs regression with Method I.  Fig. \ref{fig: db_pfpole} shows the $DB_{PF,\lambda}$ subsets, used for the PFs regression with Method II.}
    \label{fig: db_pf}
\end{figure}

The described DBs have been adequately manipulated depending on weather a SO or a MO strategy is adopted, as explained in the following and shown in Figs. \ref{fig: db_remo}-\ref{fig: db_pf}.

For the poles' real and imaginary parts regression problems, both the SO and the MO strategies have been used. Fig. \ref{fig: db_remo} shows how $\textbf{DB}_{\Re}$ has been handled in order to implement the regression with the MO strategy. The same approach has been applied to $\textbf{DB}_{\Im}$. Each $i$-th input sample $\textbf{X}_i \in \mathbb{R}^{1 \times N_f}$ is associated to an output array $\textbf{Y}_i \in \mathbb{R}^{1 \times M}$ that contains the values of the real parts of the $M$ modes. The two colors, blue and black, of the frames of the output instances indicate which samples have been used for training (blue) and testing (black) in DT's hyperparameters tuning and for splines accuracy verification. Finally, the MO poles' real and imaginary parts regression models realize the prediction of new cases using an input and obtaining an output with the same size of $\textbf{X}_i$ and $\textbf{Y}_i$.

The SO approach requires a slight DB manipulation. In order to predict $M$ real parts and $M$ imaginary parts, the same number of regressions have to be trained.  Each regression is trained using \textbf{X} as input and the real (imaginary) parts of the $m$-th pole, \textbf{$Y_{\Re(\lambda_m)}$} $\in \mathbb{R}^{N_s\times1}$ (\textbf{$Y_{\Im(\lambda_m)}$} $\in \mathbb{R}^{N_s\times1}$) as output. Fig. \ref{fig:db_reso} shows the structure of the $M$ DBs created for the poles' real part regression.

Fig. \ref{fig: db_pf} represents $\textbf{DB}_{PF}$ with 3-D structures, resulting from the concatenation of the $N_s$ simulations outcomes participating matrices. For the PFs regression only the MO strategy has been applied, but two different approaches have been used. In one case (Method I) the goal is to predict the PFs of one PV for all the poles. Then, $N$ DB subsets are generated, one for each PV ($\textbf{DB}_{Pvar}$ in Fig. \ref{fig: db_pfvar}, using different colors). Considering the DB subset associated to the $n$-th PV, for each input sample $\textbf{X}_i \in \mathbb{R}^{1\times N_f}$ the related output is an array $\textbf{Y}_{PV_{n,i}} \in \mathbb{R}^{1\times M}$, containing the PFs of $n$-ith PV, for all the poles generated by that input instance. The second approach (Method II) is aimed to predict the PFs of all the PVs for one pole. Then, $M$ DB subsets are generated, one for each pole ($\textbf{DB}_{PF,\lambda}$ in Fig. \ref{fig: db_pfpole}, using different colors). Considering the DB subset associated to the $m$-th pole, the input matrix involved, in addition to the three parameters $S_{SG share}$, $\tau_v$ and $R$, contains also the real and imaginary parts of the $m$-th pole, becoming \textbf{X} $\in \mathbb{R}^{N_s\times N_f}$, with $N_f$=5. The output related to a single input sample is an array $Y_{PF(\lambda_m),i} \in \mathbb{R}^{1\times N}$, containing the PFs of all PVs for the $m$-th pole.

Tables \ref{tab: poledb} and \ref{tab: pdfb} summarize the structures of inputs and outputs for each regression problem and for each regression strategy employed. 

\begin{table}[!ht]
%\begin{adjustwidth}{-1in}{-1in}
%Description of the structures of the DBs involved in the poles' real and imaginary parts regression. The subscript '*' refers to a generic sample.}
\label{tab: poledb}
\resizebox{\linewidth}{!}{%
\begin{tabular}{lcll}
\hline
\\[-0.7em]
\multicolumn{3}{c}{Poles' Real and Imaginary Parts  Regression}                                        \\ \\[-0.7em]  \hline \\[-0.7em] &
 \multirow{2}{*}{Regression Input} &
  \multicolumn{2}{c}{Regression Output} \\ \\[-0.7em] \cline{3-4} \\[-0.7em]
  \multicolumn{1}{c}{} & &
  \multicolumn{1}{c}{MO} &
  \multicolumn{1}{c}{SO} \\ \\[-0.7em] \hline \\[-0.7em] $\Re(\lambda)$ &
  \begin{tabular}[c]{@{}l@{}} $X_i$ $\in$ $\mathbb{R}^{1\times N_f}$\\ $X_i$=[$S_{SG share i}$,$\tau_{v i}$,$R_i$] \end{tabular}  & \begin{tabular}[c]{@{}l@{}} $Y_i$ $\in$ $\mathbb{R}^{1\times M}$\\ $Y_i$=[$\Re(\lambda_{1,i})$,...,$\Re(\lambda_{M,i})$] \end{tabular} & \begin{tabular}[c]{@{}l@{}}% $Y^*$ \subset Y; $Y*$= $\cup \big[ Y^*_{1},...,Y^*_{j},...,Y^*_{N}$\big]\\ 
 $Y_i$ $\in$ $\mathbb{R}^{1}$\\ $Y_i$=[$\Re(\lambda_{m,i})$]\\ m=1,...,M \end{tabular} \\ \\[-0.7em] \hline \\[-0.7em]
%%%%%%%%%%%%%%%%%%%%%
 $\Im(\lambda)$ & \begin{tabular}[c]{@{}l@{}} $X_i$ $\in$ $\mathbb{R}^{1\times N_f}$\\ $X_i$=[$S_{SG share i}$,$\tau_{v i}$,$R_i$] \end{tabular}  & \begin{tabular}[c]{@{}l@{}} $Y_i$ $\in$ $\mathbb{R}^{1\times M}$\\ $Y_i$=[$\Im(\lambda_{1,i})$,...,$\Im(\lambda_{M,i})$] \end{tabular} & \begin{tabular}[c]{@{}l@{}} %$Y_{SO} \subset Y$; $Y = \cup \big[Y_{SO,1},...,Y_{SO,j},...,Y_{SO,N}\big]$\\ 
 $Y_i$ $\in$ $\mathbb{R}^{1}$\\ $Y_i$= [$\Im(\lambda_{m,i})$] \\ m=1,...,M \end{tabular} \\ \\[-0.7em] \hline
\end{tabular}}
%\end{adjustwidth}
\caption{
Shape of the input and output of a $i$-th instance in the poles' real and imaginary parts regression problems.}
\end{table}

\begin{table}[!ht]
\label{tab: pdfb}
%\begin{adjustwidth}{-1in}{-1in}
\resizebox{\linewidth}{!}{%
\begin{tabular}{cccc}
%\begin{tabularx}{\linewidth}{XXX}
\hline\\[-0.7em]
\multicolumn{3}{c}{PFs Regression}   \\[-0.7em]                                       \\ \hline \\[-0.7em]
  \multirow{2}{*}{} &
  \multicolumn{1}{c}{Regression Input} &
  \multicolumn{1}{c}{Regression Output}\\[-0.7em] \\ \hline\\[-0.7em]
  \multirow{2}{*}{Method I} 
 &  \begin{tabular}[c]{@{}c@{}} $X_i$ $\in$ $\mathbb{R}^{1\times N_f}$\\ $X_i$=[$S_{SG share i}$,$\tau_{v i}$,$R_i$] \end{tabular}  & \begin{tabular}[c]{@{}c@{}} $Y_{PV {n},i}$ $\in$ $\mathbb{R}^{1 \times M}$\\ $Y_{PV{n},i}=[PF(\lambda_1),...,PF(\lambda_M)]$ \\ n=1,...,N\\ 
 %$Y_{PF,l}$ $\in$ $\mathbb{R}^{N \times Np}$
 \end{tabular} \\\\[-0.7em] \hline\\\\[-0.7em]
%%%%%%%%%%%%%%%%%%%%%
 \multirow{2}{*}{Method II} 
 &  \begin{tabular}[c]{@{}c@{}} $X_i$ $\in$ $\mathbb{R}^{1\times (N_f+2)}$\\ $X_i$=[$S_{SG share i}$,$\tau_{v i}$,$R_i$,$\Re{(\lambda_{m})_i}$,$\Im{(\lambda_{m})_i}$]\\ m=1,...,M \end{tabular}  & \begin{tabular}[c]{@{}c@{}} $Y_i$ $\in$ $\mathbb{R}^{1\times N}$\\ $Y_{PF(\lambda_m),i} \in \mathbb{R}^{1\times N}$ \\ $Y_{PF(\lambda_m),i}=[PF(\lambda_m)_{PV_1},...,PF(\lambda_m)_{PV_N}]$ \\ m=1,...,M %=[$\Im(\lambda_{n,1})$,...,$\Im(\lambda_{n,22})$] 
 \end{tabular} \\ \\[-0.7em] \hline
\end{tabular}}
%\end{adjustwidth}
\caption{Shape of the input and output of a $i$-th instance in the PFs regression problems.}
%Description of the structures of the DBs involved in the PFs regression. The subscript '*' refers to a generic sample.}
\end{table}

\subsection{Training Phase}
\label{sec: training}
\subsubsection{DT-based Regressions}
%For the regression of the real and imaginary parts both DT-based and Spline-based regressions have been implemented, with several different approaches.
DT-based regressions have been employed both for the poles and the PFs regressions.
For the poles regressions, both the SO and MO strategies have been used. When the regression is made by a single tree, the \textit{best split} has been adopted as splitting criterion and the MSE as \textit{impurity} function. The MO approach has been implemented also with a bagging ensemble technique. In this case the splitting criterion is the \textit{best random} split and the \textit{impurity} function adopted is the MSE. %In the following, SO-DT, MO-DT and ENS MO-DT refer to single-output single tree, multi-output single tree and ensemble trees multi-output models, respectively. 
Single tree single-output Decision Tree (SO-DT for short) involves the training of $2N$ models for the poles real and imaginary parts computation. With single tree multi-outputs Decision Tree (MO-DT for short) and ensemble multi-outputs Decision Tree (ENS MO-DT) only the training of two models is needed, one for the real and one for the imaginary parts.

The PFs regressions are trained only with a MO strategy, with the single tree and the bagging ensemble approaches, following Method I (MO-DT I, ENS MO-DT I) and Method II (MO-DT II, ENS MO-DT II).

When the ensemble strategy is used, the trained models are made up of fully-grown trees. Concerning the DT models obtained by the training of a single tree, the tree growth is limited by tuning the \textit{maximum depth}, \textit{minimum samples leaf} and \textit{minimum samples split} hyperparameters and by the \textit{minimal cost-complexity pruning} \cite{friedman2001elements}.  The tuning of the hyperparameters is carried out during the algorithms' validation phase, that entails the use of 80\% of the available training data for training and the remaining 20\% for testing the performances, with a cross-validation approach.

\begin{figure}[!ht]
    \centering
    \includegraphics[width=\linewidth]{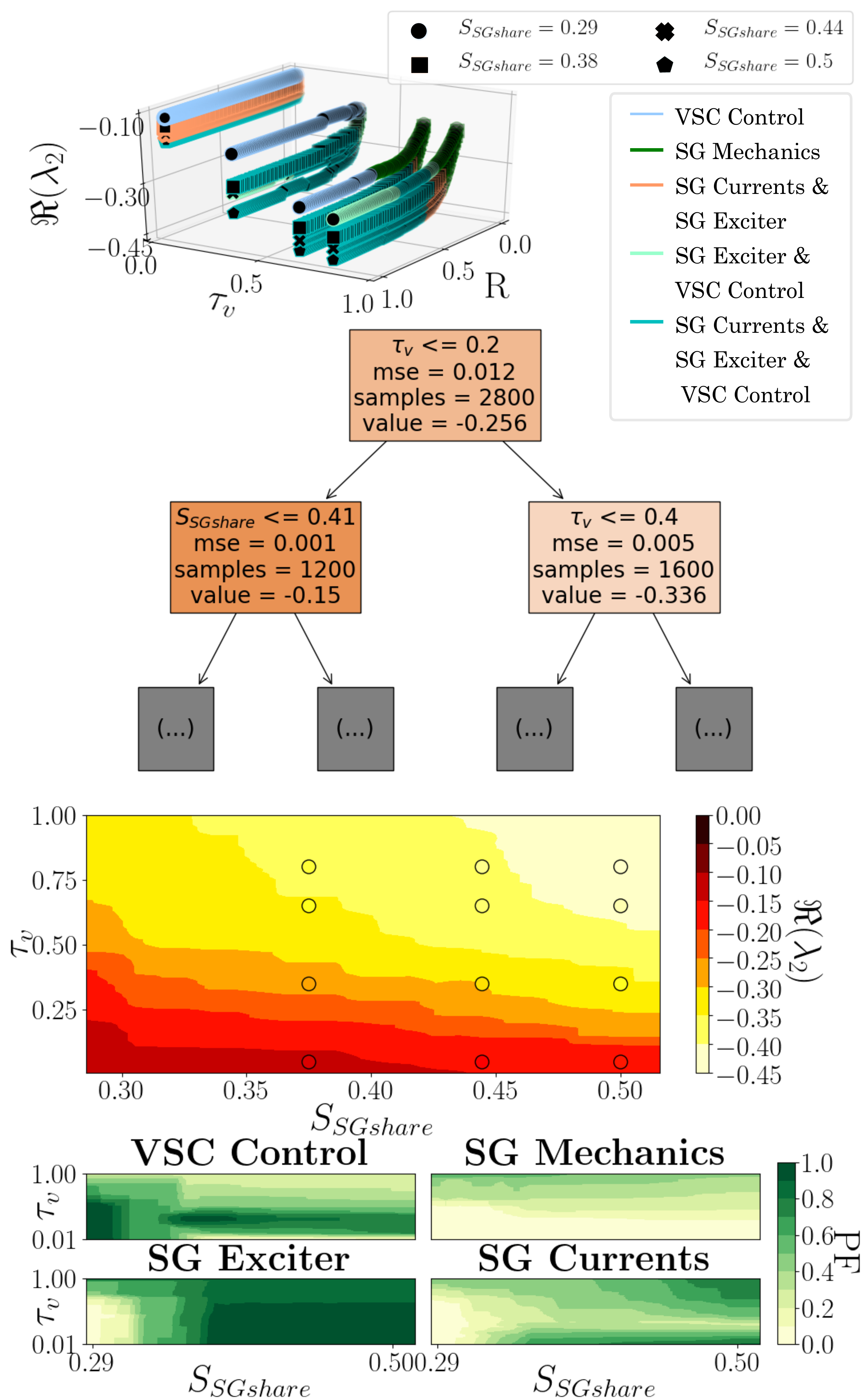}
    \caption{Example of information that it is possible to extract from a DT-based
regression model. 
Top panel: Prediction by ENS MO-DT and ENS
MO-DT Method I of a pole real part ($\Re(\lambda_2$) and of its participating variables, respectively. 
Second panel: Tree structure of the SO-DT regression model for $\Re(\lambda_2$). 
Third panel: Feature space partition of the regression (by ENS MO-DT) of $\Re(\lambda_2$), as function of the features with higher importance.
Bottom panel: Feature space partition of the regressions (by ENS MODT Method I) of the dominant participating variables groups.}
    \label{fig: dt_train}
\end{figure}

Fig. \ref{fig: dt_train} shows, by way of example, one of the trained DT models and its corresponding feature space partition representations. It refers to the computation of the real part of one of the low damped oscillation modes (namely the 2nd pole). The 3D plot at the top left panel shows the trajectories of the pole's real part,
 predicted by the trained model as explained in the following subsection. The markers edge colors correspond to the predicted PFs, which show a dominant effect of the participating variables related to the VSC Control, the SG Mechanics, the SG Exciter and the SG Currents groups. They affect the pole behaviour singularly or simultaneously. Coherently the tree structure, at the top right panel, shows that among the features considered, $S_{SG share}$ and $\tau_v$ have the highest importance, since they are involved in the splits of the upper layers.
 On the bottom of the figure, the feature space partition of different targets are 
presented. The first from above is referred to the real part of the pole, while the four at the bottom to the dominant PVs groups. Circular scatter markers in $\Re(\lambda_2)$ feature space partition show the exact value of the target for the corresponding features values, in order to have a qualitative estimation of the model accuracy.

Although the feature space partition plots are bidimensional representations, the values of the mapped target variables take also into account the effect of the other problem's features. In the shown example, following a conservative approach, the values of the mapped variables are the maximum values over the third problem's feature R.

In summary, the operational thresholds decision rules that define the tree structure are used in the feature space partition representations to map the regressions target variables. In this way a quantitative estimation of the value of the target variables is provided also for unsimulated and unknown instances. Therefore, with respect to the conventional simulation-based analysis, focused on single system conditions instances, DT trained models provide an extensive view about variables relations. This analysis improves the knowledge of the system behavior and can support the choice of possible control actions. 

\subsubsection{Spline-based Regressions}
\begin{figure}[!ht]
    \centering
    \subfloat[]{
    \includegraphics[width=\linewidth]{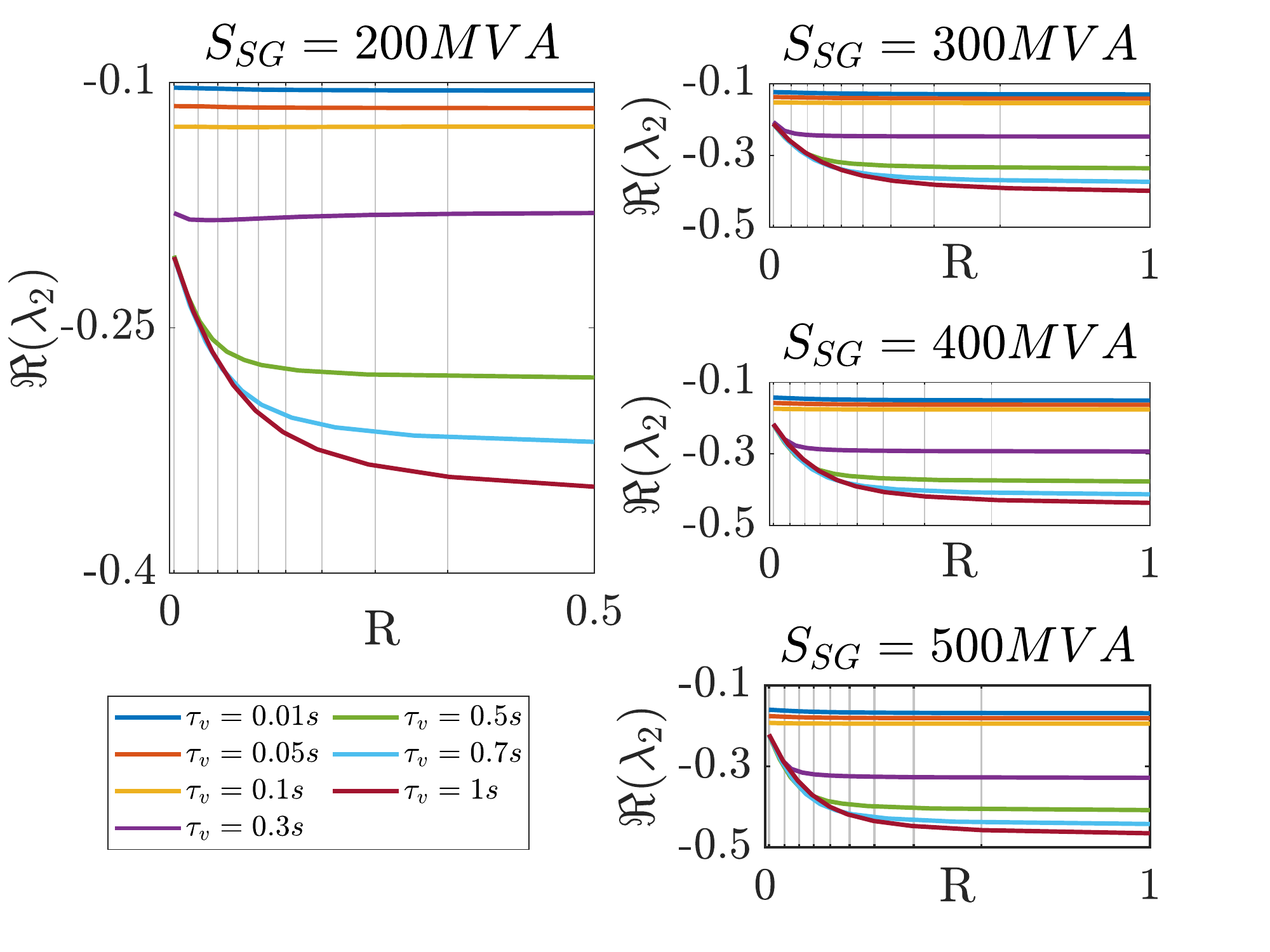}
    \label{fig: univ_spline}}
    \hfil
    %\subfloat[]{
    %\includegraphics[width=3in]{legend.PNG}
    %\label{fig: leg}}
   % \hfil
    \subfloat[]{
    \includegraphics[width=\linewidth]{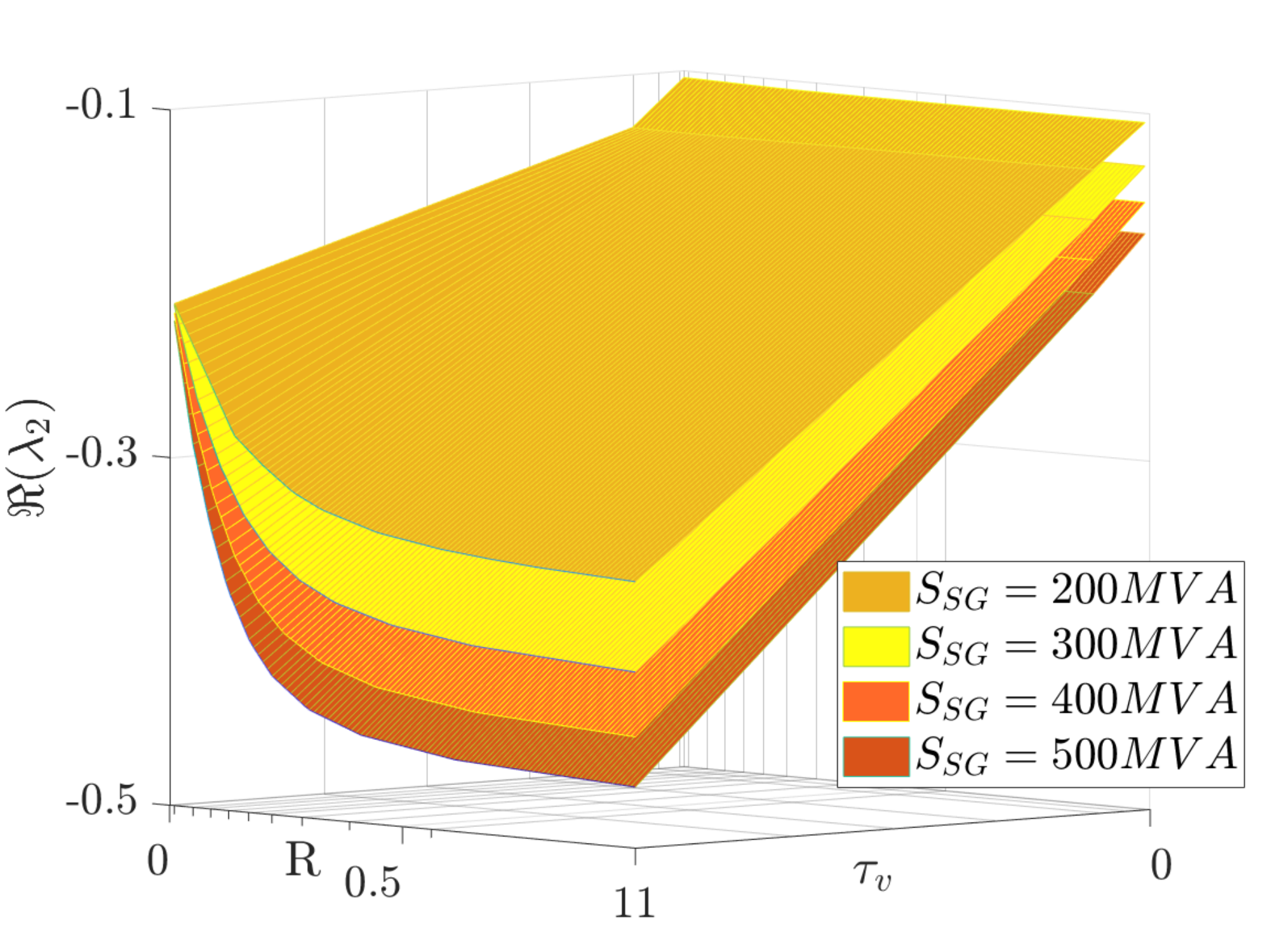}
    \label{fig: biv_spl}}
    \caption{Example of computed splines, related to the real part of the 2-nd pole. Fig. \ref{fig: univ_spline} shows the curves obtained with univariate spline approximations, Fig. \ref{fig: biv_spl} shows the surfaces obtained with  bivariate spline approximations.}
    \label{fig: spline}
\end{figure}

Spline-based regressions are employed only in the poles’ real and imaginary parts regression problems. 
The univariate interpolations (1DLI) and approximations (1DLA) are developed as function of the feature with the largest cardinality, i.e. $R$. Thus, for each combination of the values of the other features, a curve is computed, with the interpolation or the approximation method, through the estimation of the spline coefficients and knots.
The bivariate approximations (2DLA) are developed as function of $R$ and $\tau_v$: for each value of the $S_{SG share}$ present in the training DB a surface is computed. % The knots sequence adopted for the basic interval along the $R$ axis is extracted by the univariate approximation analysis. 
Fig. \ref{fig: spline} shows, as example, the curves and the surfaces obtained by the 1DLA and 2DLA of the real part of the 2nd pole. While the interpolation method fits the original curves passing through the training points, the approximation generates least square approximation curves passing through the points corresponding to the optimal knots sequence. The optimal knots sequence along the $R$ axis obtained for the univariate splines has been used also for the bivariate approximation. In Fig. \ref{fig: univ_spline}-\ref{fig: biv_spl} the average position of the break points is indicated by the grid lines. 

\subsection{Testing phase}
\label{subsec: testingph}

The testing phase consists in the deployment of the trained models for the computation of outputs corresponding to instances not belonging to the training DBs.

For DT-based algorithms, the computation of a new instance simply corresponds to the evaluation of the outcome at the end of the tree's path. The proper path is obtained by the application of the operational threshold rules, defined during training.

In the case of the splines regression, a multiple linear interpolation approach is used. Consider a generic new input instance $X_*=[S_{SG share *}, \tau_{v*}, R_*]$. Among all the splines curves (or surfaces), the ones involved in the computation are the splines related to values of the input features that are the lower and upper values closest to the actual input. They are used to compute the values of the target variable as function of $R_*$ or of ($R_*,\tau_{v*})$  whether univariate and bivariate splines are adopted. Then, between the obtained values, simple linear interpolations are applied to evaluate the final output prediction. 

In order to compare the performances of the models, the trained regressions are tested with the following features values: $S_{SG share}$ = 330 MVA, $\tau_v$=[0.05, 0.2, 0.35, 0.5, 0.65, 0.8, 0.95] s and $R$=0.05. The performances are evaluated for several values of $\tau_v$ in order to check if the models are able to generate predictions with a good accuracy for those input features that have low cardinality in the training DB.

\begin{figure}[!ht]
    \centering
    \includegraphics[width=\linewidth]{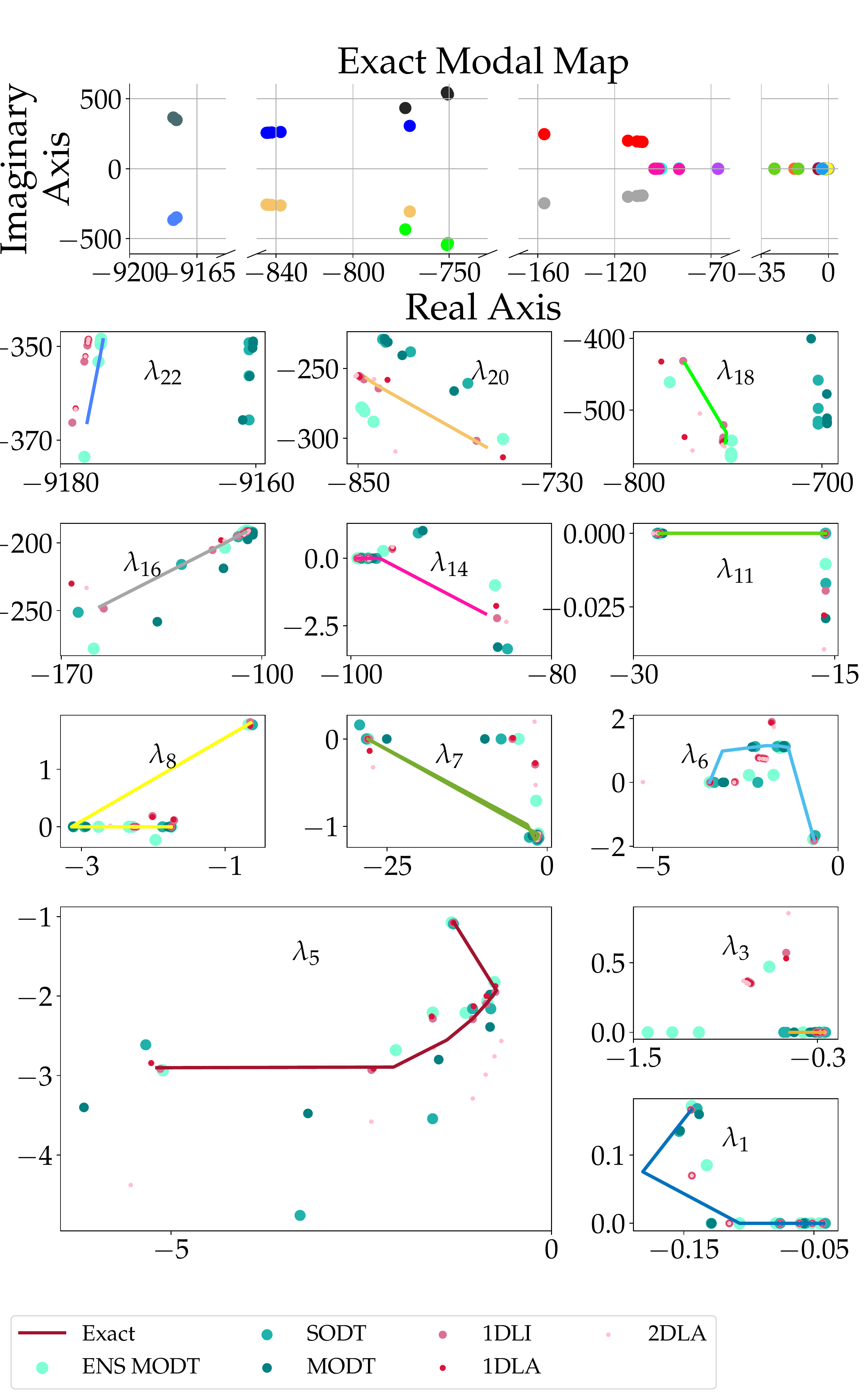}
    \caption{Comparison between exact and predicted poles, obtained for the instances: $S_{SG share}$ = 330 MVA, $\tau_v$=[0.05, 0.2, 0.35, 0.5, 0.65, 0.8, 0.95]s and $R$=0.05.}
    \label{fig: tvres}
\end{figure}

Table \ref{tab: respoletv} summarizes the performances achieved by all the techniques tested for the regression of the poles’ real and imaginary parts. Concerning the training and testing computing times, consider that both time domain simulations and data-driven computation have been executed with a CPU Intel Core i5, 8 GB of RAM. The most accurate methods are the DT-based ones and, among them, the ENS MO-DT is the best. Concerning the splines-based regressions, the most accurate is 1DLA, which error is slightly lower than the 1DLI. The accuracy is measured using the Wave Hedge Error (WHE), for a scale free error estimation  \cite{botchkarev2018performance}. Fig. \ref{fig: tvres} compares the modal maps obtained by all the regression techniques with the exact values, obtained by simulations. The top panel shows the exaxt modal map and identifies the \textit{M} poles thanks to the use of different marker colors. According to the colors used in the plot of the exact modal map, the remaining subplots show the exact (solid lines) and predicted (circular markers) location of some selected poles.

Fig. \ref{fig: tvres} confirms the accuracy results summarized in Table \ref{tab: respoletv} and allows to estimate the different algorithms' generalization ability. Considering the 5-th pole ($\lambda_5$) by way of example, it is possible to notice that, among the DT-based regressions, only the ENS MO-DT is able to compute seven distinct solutions, one for each test instance. SO-DT and MO-DT, are able to compute only five and six distinct solutions, respectively. In fact, using MO-DT, the solutions for $\tau_v$ equal to 0.65 s and 0.8 s coincide and are equal to the one that would be obtained with $\tau_v$ equal to 0.7 s, which is contained in the training DB. This demonstrate that SO-DT and MO-DT require a larger training DB and that, with the same training DB, the ensemble method has a larger generalization ability. 

\begin{table}[]
\label{tab: respoletv}
\resizebox{\linewidth}{!}{
\begin{tabular}{ccccc}
\hline
\multicolumn{5}{c}{Poles' Real and Imaginary parts Regression}         \\ \hline
\multirow{2}{*}{Method} & \multirow{2}{*}{\begin{tabular}[c]{@{}c@{}}Training\\ time {[}s{]}\end{tabular}} & \multirow{2}{*}{\begin{tabular}[c]{@{}c@{}}Testing\\ time {[}s{]}\end{tabular}} & \multicolumn{2}{c}{Average Error} \\ \cline{4-5} 
          &   &  & $\Re$     &  $\Im$     \\ \hline
SO-DT     & 0.1348 & 0.008  & 0.0472 & 0.151 \\
MO-DT     & 0.0440 & 0.0006 & 0.0552 & 0.0767 \\
\bf{ENS MO-DT} & \bf{0.2015} & \bf{0.0014} & \bf{0.0467} & \bf{0.1117} \\
1DLI      & 3.5348 & 0.0264 & 0.0848 & 0.2107 \\
1DLA      & 4.3997 & 0.0330 & 0.0491 & 0.2137 \\
2DLA      & 3.6521 & 0.0374 & 0.0540 & 0.2332 \\ \hline
\end{tabular}
}
\caption{Performances comparison of the trained models relative to the poles' real and imaginary parts regression problem.}
\end{table}

\begin{figure}[!ht]
    \centering
    \includegraphics[width=\linewidth]{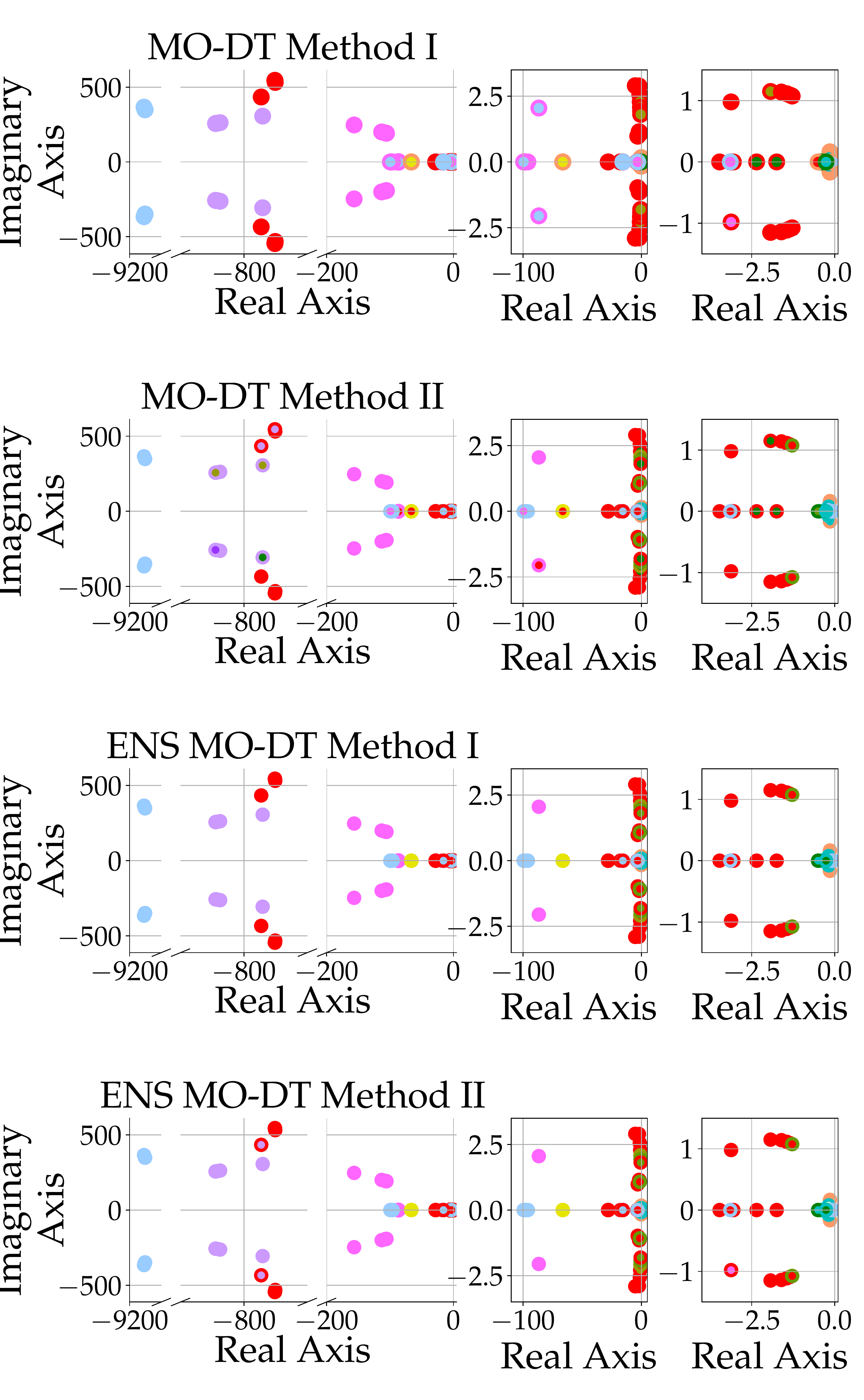}
    \caption{Comparison between exact PFs (indicated by the markers edge color) and predicted PFs (indicated by the markers inner color).}
    \label{fig: pfres}
\end{figure}

Table \ref{tab: pfres} summarizes the performances achieved for the PFs' regression problem. The error is estimated as the MAE.  In Fig. \ref{fig: pfres} the modal maps of the test cases are shown. The colors of the poles indicate which controllers have a dominant impact on the pole. %, following the legend table in \ref{}.
In particular, for each pole marker the edge color indicates the dominant impact obtained with the exact PFs, while the inner color indicates the dominant impact corresponding to the predicted PFs. If the poles show different edge and inner color, means that the error of the regression brings to a misclassification of the dominant controllers.  

\begin{table}[]
\label{tab: pfres}
\resizebox{\linewidth}{!}{%
\begin{tabular}{ccccc}
\hline \\[-0.7em]
\multicolumn{5}{c}{PFs' Regression}                                              \\[-0.7em]     \\ \hline \\[-0.7em]
Method      & \begin{tabular}[c]{@{}c@{}}Training\\ time {[}s{]}\end{tabular} & \begin{tabular}[c]{@{}c@{}}Testing\\ time {[}s{]}\end{tabular} & \begin{tabular}[c]{@{}c@{}}Average\\ error \end{tabular} & \begin{tabular}[c]{@{}c@{}}Misclassified\\ cases\end{tabular} \\[-0.7em] \\ \hline \\[-0.7em]
MO-DT 1     & 1.8098            & 0.0635        & 0.1102        & 32.5\%              \\
MO-DT 2     & 1.7591            & 0.1381        & 0.0949        & 34.4\%              \\
\bf ENS MO-DT 1 & \bf 1.4047 & \bf 0.2916 & \bf 0.0221 & \bf 7.1\% \\
ENS MO-DT 2 & 2.639             & 0.2903        & 0.0280        & 14.3\%     \\[-0.7em]         \\ \hline
\end{tabular}%
}
\caption{Performances comparison of the trained models relative to the PFs regression problem.}
\end{table}

\begin{figure}[!ht]
    \centering
    \includegraphics[width=\linewidth]{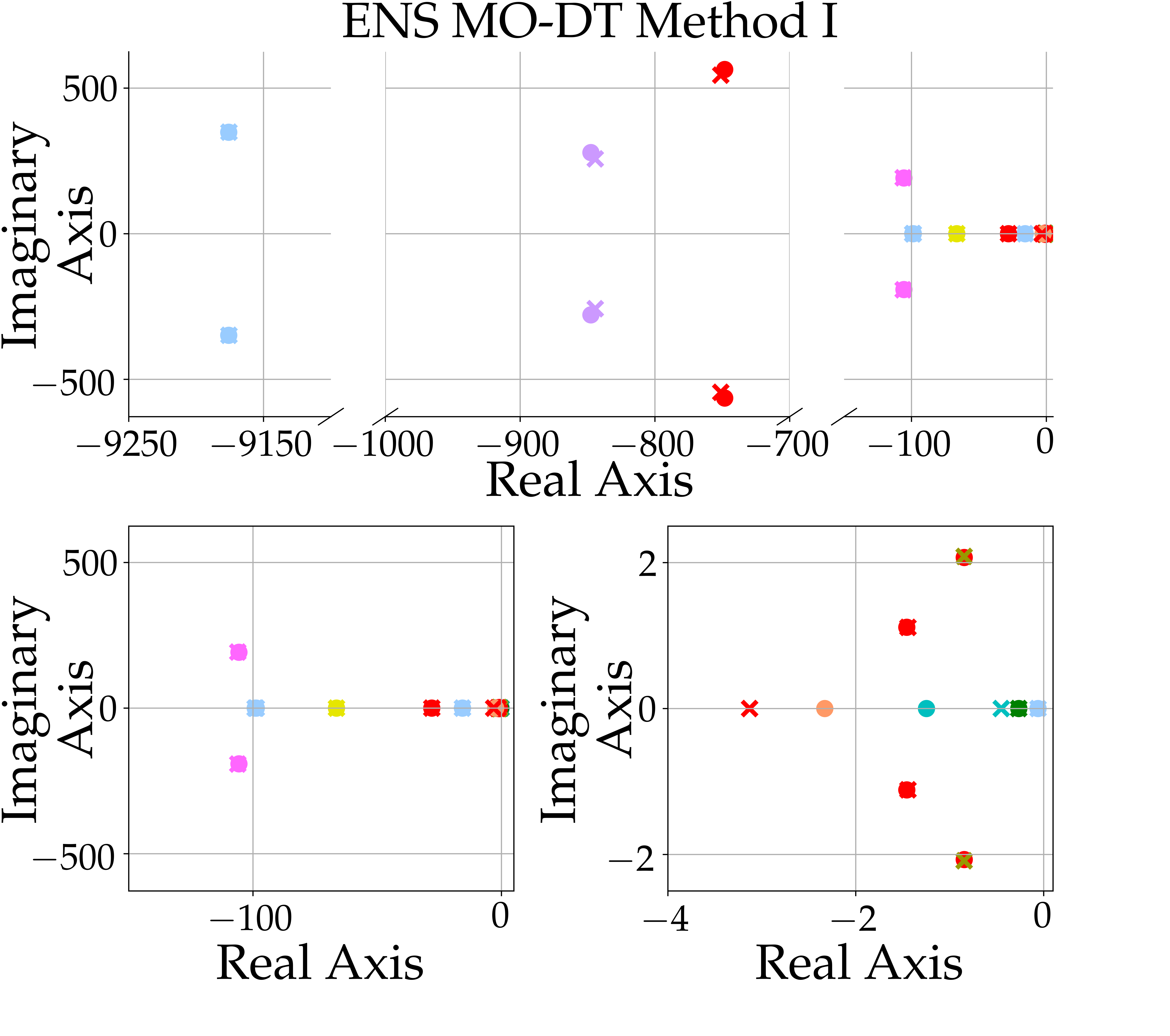}
    \caption{Comparison of the exact (cross-shaped markers) and predicted (circle shaped
markers) poles location. The color of the markers indicates the exact and predicted resulting dominant participating variable groups, according to their PFs. (Test instance: $S_{SG share}=330$ MVA, $\tau_v=0.65$ s, $R=0.05$)}
    \label{fig: res_fin}
\end{figure}

Finally, the solution of the overall problem (regression of poles' real and imaginary parts and regression of corresponding PFs), for a single test case, is compared with a single simulation result. For $S_{SG share}$=330 MVA, $\tau_v$=0.65 s, $R$=0.05 the time-domain non-linear model, followed by the state-space model and the eigenvalues analysis takes 102 s. The ensemble DTs take 0.9253 s with an error equal to 0.0429 and to 0.0113 for the real and imaginary parts, respectively. The error over the prediction of the PFs is equal to 0.0123, that brings to the misclassification of only three poles (Fig. \ref{fig: res_fin}).

\subsection{Test on a larger power system}
\label{subsec: testcase2}
\begin{figure}[!ht]
    \centering
    \includegraphics[width=\linewidth]{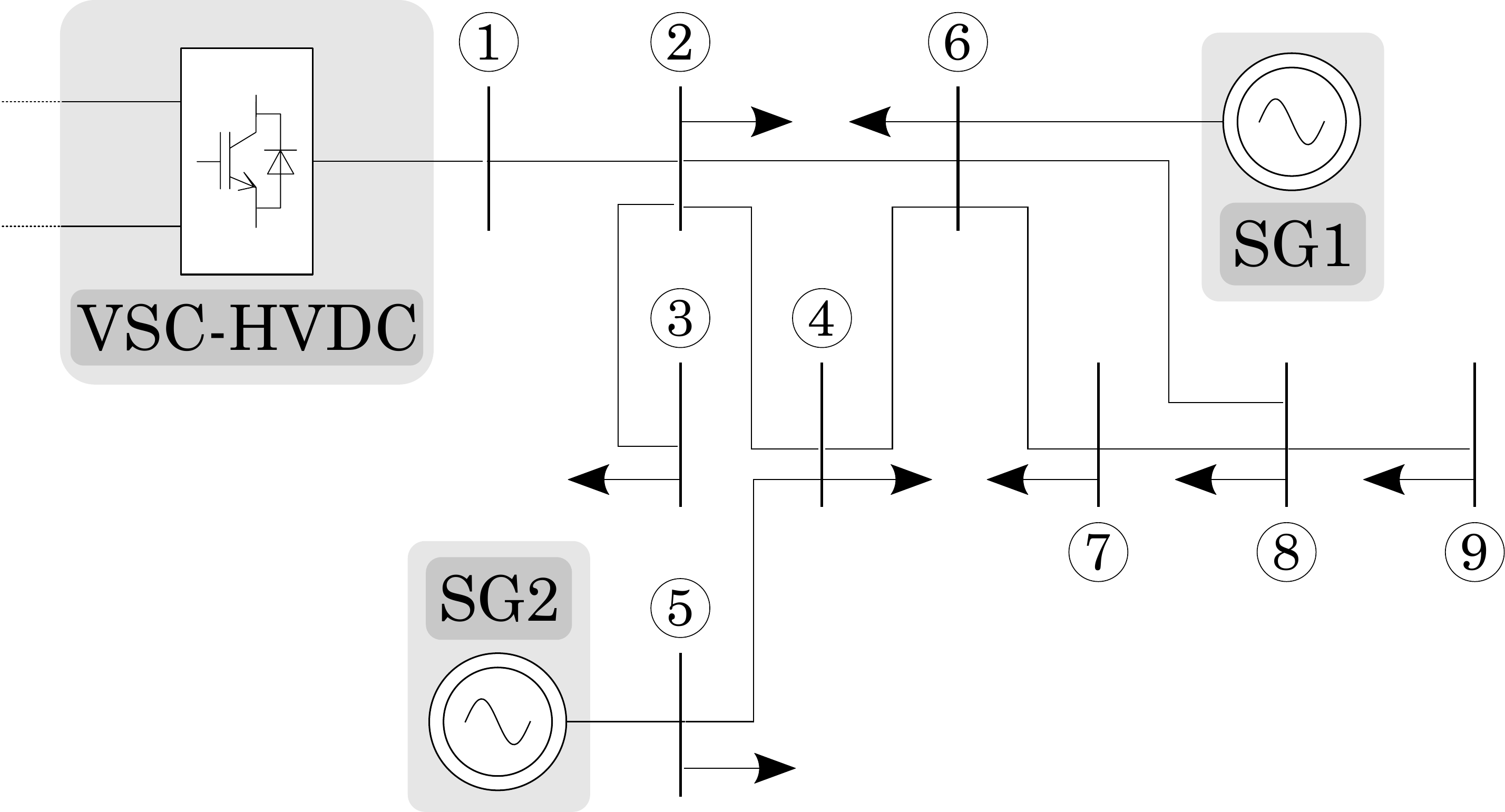}
    \caption{Scheme of the 9-bus system}
    \label{fig: syst2}
\end{figure}

The proposed data driven methodology has been tested on a larger power system model. The scheme of the system used is shown in Fig. \ref{fig: syst2}. It is still an essential model of an converter-based power system, made up of nine buses with eight loads, two SGs and one VSC, representing power demand and the aggregated thermal and renewable based generation. The SGs rated power has been progressively decreased while the VSC rated power is fixed to 500 MVA. 

Data generation process involves power flows, state space linearization and eigenvalues 
analysis. The simulations have been executed considering that: %The total amount of simulations executed is $N_s=5\times5\times3\times7\times7$:
\begin{itemize}
    \item Each SG rated power is progressively decreased from 250 MVA to 50 MVA, with a step of 50 MVA;
    \item Three power demand scenarios are considered: First the total demand, distributed among the eight loads, is set equal to 500 MW; Then, 10\% of power demand increase is simulated, changing the power absorbed at loads 5 (from 100 MW to 150 MW) and 6 (from 50 MW to 100 MW);
    \item The VSC voltage controller time constant $\tau_v$ is set equal to 0.01s, 0.05s, 0.1s, 0.3s, 0.5s,0.7s, 1s;
    \item The VSC frequency droop characteristic is set equal to 0.01, 0.05, 0.1, 0.3, 0.5,0.7, 1.
\end{itemize}

Therefore, the total amount of simulations is $N_s= 5\times5\times3\times7\times7= 3675$ and $N_f= 6$ features are considered: $S_{SG1 share}$, $S_{SG2 share}$, $P_{load 5, share}$, $P_{load 6, share}$, $tau_v$, $R$. The state space analysis involves $N=71$ state variables. Thus, the system response to each simulation instance results in $M=N=71$ oscillation modes. Therefore, a generic $i$-th simulation generates outputs with the following shapes: $Y_{\Re,i} \in \mathbb{R}^{1\times M}$, $Y_{\Im,i} \in \mathbb{R}^{1\times M}$, $Y_{PF,i} \in \mathbb{R}^{N\times M}$. 

The regression technique implemented both for the poles and PFs regressions is the one that achieved the best performances in the previous test case, namely the ENS MO-DT. According to this choice, the DBs have been arranged following the methodology described in Section \ref{sec: dborg} for the application of a MO strategy.

\begin{figure}[!ht]
    \centering
    \includegraphics[width=\linewidth]{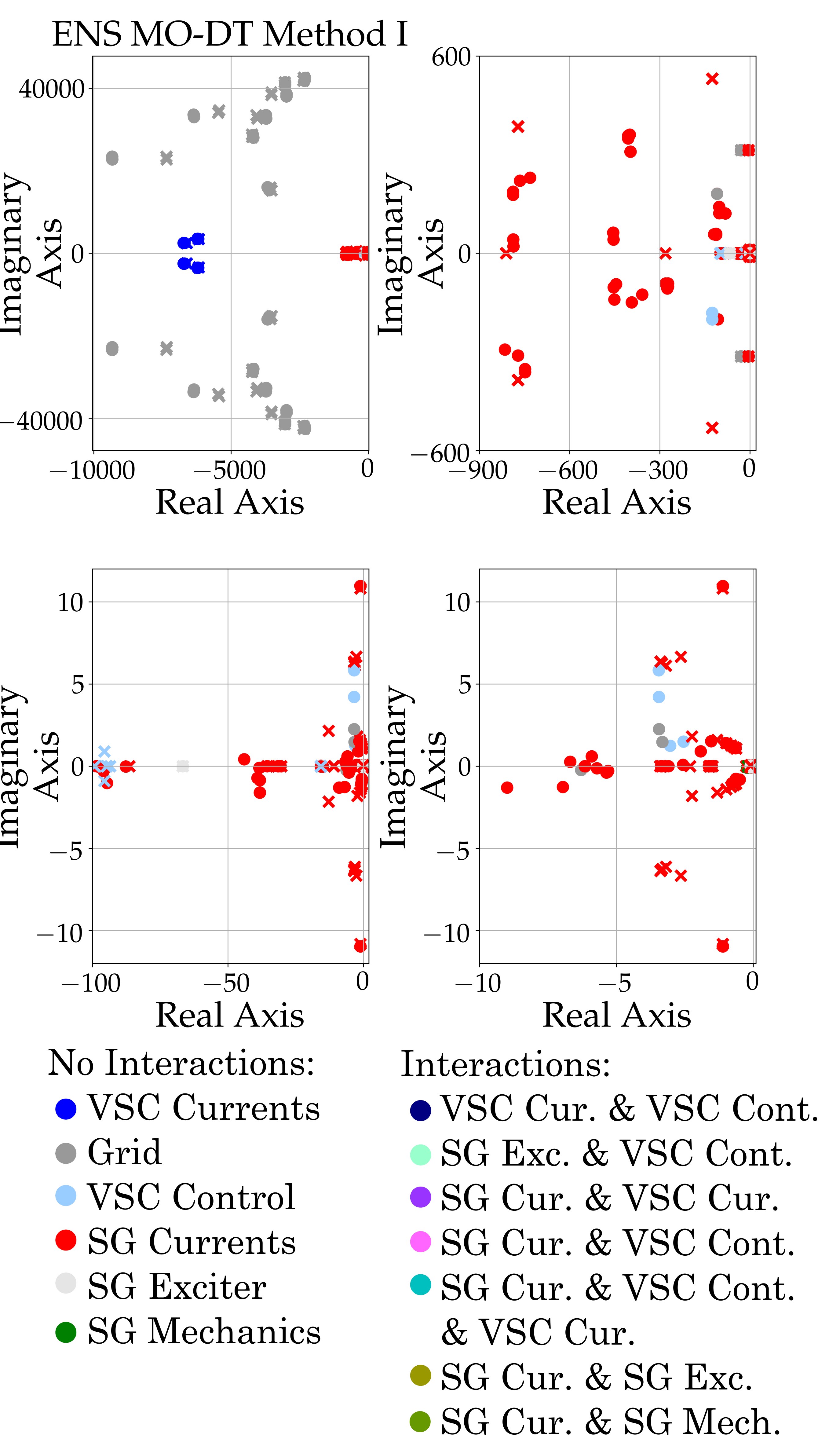}
    \caption{Comparison between the exact (cross-shaped markers) and predicted (circle-shaped markers) poles location. The color of the markers indicates the exact and predicted resulting dominant participating variable groups, according to their PFs and coherently with the legend below the plots.}
    \label{fig: res_syst2}
\end{figure}

Table \ref{tab: res2} summarizes the results obtained for the following test cases: $S_{SG1 share}= 0.28$ ($S_{SG1}= 70$ MVA), $S_{SG2 share}= 0.68$ ($S_{SG2}= 170$ MVA), $P_{load 5, share}= 0.5$ ($P_{load 5}= 125 $MW), $P_{load 6, share}= 0.28$ ($P_{load 6}=70$ MW), $\tau_v= [0.05,0.2,0.35,0.5,0.65,0.8,0.95]$s, $R= 0.02$. For the same test cases, Fig. \ref{fig: res_syst2} shows a comparison between the exact (cross-shaped markers) and predicted (circle-shaped markers) modal maps. The colors of the markers indicate the dominant participating variables groups, following the legend below the plots.

\begin{table}[]
\label{tab: res2}
\begin{tabular}{cccc}%x}{XXXX}
\hline
\multicolumn{4}{c}{Regression}                                                                   \\ \hline
\multicolumn{1}{l}{}                    & \multicolumn{2}{c}{Poles}       & \multirow{2}{*}{PFs} \\
\multicolumn{1}{l}{}                    & $\Re(\lambda)$ & $\Im(\lambda)$ &                      \\ \hline
Training time [s]                       & \multicolumn{2}{c}{1.2001}      & 21.5                 \\
Test time [s]                           & \multicolumn{2}{c}{0.0022}      & 3.1                  \\
Average error                           & 0.1162         & 0.1925       & 0.0361               \\
\multicolumn{1}{l}{Misclassified cases} & -              & -              &  15.9\%                \\ \hline
\end{tabular}%x}
\caption{Performances achieved by ENS MO-DT and ENS MO-DT Method I for the regression of the modal map and the PFs of the larger power system test case, respectively.}
\end{table}

\section{ Conclusion and discussion on data-driven model applications}
\label{sec: conclusion}
In this study a complete methodology for a data-driven small-signal power system stability analysis has been developed, based on the implementation of regression techniques. DT-based and Spline based regression have been implemented for the prediction of the 
modal map of the system and of the PFs. The methodology has been tested on two essential power systems, with the characteristics of converter-based systems and with an increasing level of complexity. The regression models have been trained using as features physic variables, as the SGs rated power, also related to the system operation point, as the power demand, and converter control variables. However, as discussed in Section \ref{subsec: new_meth}, the feature selection could be also extended to other variables. 
Among the trained model, the DT-based regressions perform with highest accuracy and lowest training and testing computing time. In particular, the use of the ensemble strategy improves the generalization ability of the models. A dramatic reduction in the computing time is achieved, if compared to the use of conventional
time-domain simulations. Besides, the ability of DT regression to realize feature space partition can be used to map the target of the regression as function of the selected features, providing a fast and broad view on the system behaviour.

%The applications of the data-driven model derived are many and varied. This section includes 
In the following, a list of the potential uses of the derived model is provided:
\begin{itemize}

\item Extraction of additional information and mapping of small-signal stability results: %DT model provides in a fast manner a wide overview of the system behavior as function of the selected features, 
find out the cross-relations among variables that might not be evident employing conventional small-signal analysis techniques%, which provide information only about the single test instance simulated.
%\item Interpolation and extrapolation. Perform the small signal stability study for new operating point, without executing the real model again, simply evaluating the data-driven obtained functions. Regressions model able to perform extrapolation could be investigated as well as strategies for the updating of the training DB and for the retraininig of the model, aimed to expand the feasible region of the model. 
\item Data-driven IP protected model: provision of a closed model based on equations revealing the dynamics of the system without providing the internal details, for network studies. 
\item System optimization: use the data-driven model in order to include dynamic constraints in system optimizations for either planning and operation analysis.
\item Extract stability indicators:  use the data-driven small-signal estimation to compute stability indicators and build a system stability margin in a low intensive computation manner.

 %   \item Extrapolation and interpolation uses. Find new operation points without the need of executing the real model again, simply evaluating the data-driven obtained functions. Regressions could be also used in order to provide information by means of extrapolation, in order to find potential issues, e.g. instabñle operation points, among others.
\end{itemize}

 \bibliographystyle{elsarticle-num} 
\bibliography{bibliography}

\begin{thebibliography}{10}
\expandafter\ifx\csname url\endcsname\relax
  \def\url#1{\texttt{#1}}\fi
\expandafter\ifx\csname urlprefix\endcsname\relax\def\urlprefix{URL }\fi
\expandafter\ifx\csname href\endcsname\relax
  \def\href#1#2{#2} \def\path#1{#1}\fi

\bibitem{vom2020data}
F.~vom Scheidt, H.~Medinov{\'a}, N.~Ludwig, B.~Richter, P.~Staudt,
  C.~Weinhardt, Data analytics in the electricity sector--a quantitative and
  qualitative literature review, Energy and AI 1 (2020) 100009.
\newblock \href {https://doi.org/https://doi.org/10.1016/j.egyai.2020.100009}
  {\path{doi:https://doi.org/10.1016/j.egyai.2020.100009}}.

\bibitem{entso2015tso}
C.~ENTSO-E, E.~GEODE, G.~EFS, Tso-dso data management report, Brussels, TSO DSO
  data management Project Team, Tech. Rep (2015).

\bibitem{hirth2018entso}
L.~Hirth, J.~M{\"u}hlenpfordt, M.~Bulkeley, The entso-e transparency
  platform--a review of europe’s most ambitious electricity data platform,
  Applied energy 225 (2018) 1054--1067.
\newblock \href
  {https://doi.org/https://doi.org/10.1016/j.apenergy.2018.04.048}
  {\path{doi:https://doi.org/10.1016/j.apenergy.2018.04.048}}.

\bibitem{hirth2020open}
L.~Hirth, Open data for electricity modeling: Legal aspects, Energy Strategy
  Reviews 27 (2020) 100433.
\newblock \href {https://doi.org/https://doi.org/10.1016/j.esr.2019.100433}
  {\path{doi:https://doi.org/10.1016/j.esr.2019.100433}}.

\bibitem{idehen2019pmu}
I.~Idehen, W.~Jang, T.~Overbye, Pmu data feature considerations for realistic,
  synthetic data generation, in: 2019 North American Power Symposium (NAPS),
  IEEE, 2019, pp. 1--6.
\newblock \href {https://doi.org/10.1109/NAPS46351.2019.9000335}
  {\path{doi:10.1109/NAPS46351.2019.9000335}}.

\bibitem{hatziargyriou2020stability}
N.~Hatziargyriou, J.~Milanovi{\'c}, C.~Rahmann, V.~Ajjarapu, C.~Ca{\~n}izares,
  I.~Erlich, D.~Hill, I.~Hiskens, I.~Kamwa, B.~Pal, et~al., Stability
  definitions and characterization of dynamic behavior in systems with high
  penetration of power electronic interfaced technologies (2020).
\newblock \href {https://doi.org/http://hdl.handle.net/2268/247834}
  {\path{doi:http://hdl.handle.net/2268/247834}}.

\bibitem{shah2015review}
R.~Shah, N.~Mithulananthan, R.~Bansal, V.~Ramachandaramurthy, A review of key
  power system stability challenges for large-scale pv integration, Renewable
  and Sustainable Energy Reviews 41 (2015) 1423--1436.
\newblock \href {https://doi.org/https://doi.org/10.1016/j.rser.2014.09.027}
  {\path{doi:https://doi.org/10.1016/j.rser.2014.09.027}}.

\bibitem{6513320}
S.~Eftekharnejad, V.~Vittal, G.~T. Heydt, B.~Keel, J.~Loehr, Small signal
  stability assessment of power systems with increased penetration of
  photovoltaic generation: A case study, IEEE Transactions on Sustainable
  Energy 4~(4) (2013) 960--967.
\newblock \href {https://doi.org/10.1109/TSTE.2013.2259602}
  {\path{doi:10.1109/TSTE.2013.2259602}}.

\bibitem{he2013small}
P.~He, F.~Wen, G.~Ledwich, Y.~Xue, Small signal stability analysis of power
  systems with high penetration of wind power, Journal of Modern Power Systems
  and Clean Energy 1~(3) (2013) 237--244.
\newblock \href {https://doi.org/10.1007/s40565-013-0028-9}
  {\path{doi:10.1007/s40565-013-0028-9}}.

\bibitem{feilat2007neural}
E.~Feilat, Neural network based assessment of small-signal stability,
  International Journal of Modelling and Simulation 27~(2) (2007) 151--157.
\newblock \href
  {https://doi.org/https://doi.org/10.1080/02286203.2007.11442411}
  {\path{doi:https://doi.org/10.1080/02286203.2007.11442411}}.

\bibitem{teeuwsen2003small}
S.~Teeuwsen, I.~Erlich, M.~El-Sharkawi, Small-signal stability assessment based
  on advanced neural network methods, in: 2003 IEEE Power Engineering Society
  General Meeting (IEEE Cat. No. 03CH37491), Vol.~4, IEEE, 2003, pp.
  2349--2649.
\newblock \href {https://doi.org/10.1109/PES.2003.1271000}
  {\path{doi:10.1109/PES.2003.1271000}}.

\bibitem{cao2018deep}
J.~Cao, Z.~Fan, Deep learning-based online small signal stability assessment of
  power systems with renewable generation, in: 2018 IEEE SmartWorld, Ubiquitous
  Intelligence \& Computing, Advanced \& Trusted Computing, Scalable Computing
  \& Communications, Cloud \& Big Data Computing, Internet of People and Smart
  City Innovation (SmartWorld/SCALCOM/UIC/ATC/CBDCom/IOP/SCI), IEEE, 2018, pp.
  216--221.
\newblock \href {https://doi.org/10.1109/SmartWorld.2018.00072}
  {\path{doi:10.1109/SmartWorld.2018.00072}}.

\bibitem{fu2020data}
Y.~Fu, L.~Chen, Z.~Yu, Y.~Wang, D.~Shi, Data-driven low frequency oscillation
  mode identification and preventive control strategy based on gradient
  descent, Electric Power Systems Research 189 (2020) 106544.
\newblock \href {https://doi.org/https://doi.org/10.1016/j.epsr.2020.106544}
  {\path{doi:https://doi.org/10.1016/j.epsr.2020.106544}}.

\bibitem{teeuwsen2005genetic}
S.~Teeuwsen, I.~Erlich, M.~El-Sharkawi, U.~Bachmann, Genetic algorithm and
  decision tree based oscillatory stability assessment, in: 2005 IEEE Russia
  Power Tech, IEEE, 2005, pp. 1--7.
\newblock \href {https://doi.org/10.1109/TPWRS.2006.873408}
  {\path{doi:10.1109/TPWRS.2006.873408}}.

\bibitem{wehenkel1989artificial}
L.~Wehenkel, T.~Van~Cutsem, M.~Ribbens-Pavella, An artificial intelligence
  framework for online transient stability assessment of power systems, IEEE
  Transactions on Power Systems 4~(2) (1989) 789--800.
\newblock \href {https://doi.org/10.1109/59.193853}
  {\path{doi:10.1109/59.193853}}.

\bibitem{wehenkel1994decision}
L.~Wehenkel, M.~Pavella, E.~Euxibie, B.~Heilbronn, Decision tree based
  transient stability method a case study, IEEE Transactions on Power Systems
  9~(1) (1994) 459--469.
\newblock \href {https://doi.org/10.1109/59.317577}
  {\path{doi:10.1109/59.317577}}.

\bibitem{zheng2012regression}
C.~Zheng, V.~Malbasa, M.~Kezunovic, Regression tree for stability margin
  prediction using synchrophasor measurements, IEEE Transactions on Power
  Systems 28~(2) (2012) 1978--1987.
\newblock \href {https://doi.org/10.1109/TPWRS.2012.2220988}
  {\path{doi:10.1109/TPWRS.2012.2220988}}.

\bibitem{thams2017data}
F.~Thams, L.~Halilbasic, P.~Pinson, S.~Chatzivasileiadis, R.~Eriksson,
  Data-driven security-constrained opf, in: Proc. 10th Bulk Power Syst. Dyn.
  Control Symp., 2017, pp. 1--10.

\bibitem{venzke2020neural}
A.~Venzke, D.~T. Viola, J.~Mermet-Guyennet, G.~S. Misyris,
  S.~Chatzivasileiadis, \href{arXiv:2003.07939}{Neural networks for encoding
  dynamic security-constrained optimal power flow to mixed-integer linear
  programs}, arXiv preprint arXiv:2003.07939 (2020).
\newline\urlprefix\url{arXiv:2003.07939}

\bibitem{kundur2007power}
P.~Kundur, Power system stability, Power system stability and control (2007)
  7--1.

\bibitem{perez1982selective}
I.~J. Perez-Arriaga, G.~C. Verghese, F.~C. Schweppe, Selective modal analysis
  with applications to electric power systems, part i: Heuristic introduction,
  ieee transactions on power apparatus and systems (1982) 3117--3125\href
  {https://doi.org/10.1109/TPAS.1982.317524}
  {\path{doi:10.1109/TPAS.1982.317524}}.

\bibitem{friedman2001elements}
J.~Friedman, T.~Hastie, R.~Tibshirani, et~al., The elements of statistical
  learning, Vol.~1, Springer series in statistics New York, 2001.

\bibitem{kamwa2011accuracy}
I.~Kamwa, S.~Samantaray, G.~Jo{\'o}s, On the accuracy versus transparency
  trade-off of data-mining models for fast-response pmu-based catastrophe
  predictors, IEEE Transactions on Smart Grid 3~(1) (2011) 152--161.
\newblock \href {https://doi.org/10.1109/PESMG.2013.6672548}
  {\path{doi:10.1109/PESMG.2013.6672548}}.

\bibitem{carvalho2019machine}
D.~V. Carvalho, E.~M. Pereira, J.~S. Cardoso, Machine learning
  interpretability: A survey on methods and metrics, Electronics 8~(8) (2019)
  832.
\newblock \href {https://doi.org/https://doi.org/10.3390/electronics8080832}
  {\path{doi:https://doi.org/10.3390/electronics8080832}}.

\bibitem{scikit-learn}
F.~Pedregosa, G.~Varoquaux, A.~Gramfort, V.~Michel, B.~Thirion, O.~Grisel,
  M.~Blondel, P.~Prettenhofer, R.~Weiss, V.~Dubourg, J.~Vanderplas, A.~Passos,
  D.~Cournapeau, M.~Brucher, M.~Perrot, E.~Duchesnay, Scikit-learn: Machine
  learning in {P}ython, Journal of Machine Learning Research 12 (2011)
  2825--2830.

\bibitem{breiman1984friedman}
O.~S. Breiman, Friedman, classification and regression trees, brooks (1984).

\bibitem{ernst2005tree}
D.~Ernst, P.~Geurts, L.~Wehenkel, Tree-based batch mode reinforcement learning,
  Journal of Machine Learning Research 6 (2005) 503--556.

\bibitem{aluja2003stability}
T.~Aluja-Banet, E.~Nafria, Stability and scalability in decision trees,
  Computational Statistics 18~(3) (2003) 505--520.
\newblock \href {https://doi.org/https://doi.org/10.1007/BF03354613}
  {\path{doi:https://doi.org/10.1007/BF03354613}}.

\bibitem{mat_spl_toolbox}
F.~Martin (2021).
\newblock
  \href{https://www.mathworks.com/matlabcentral/fileexchange/72654-spline-toolbox}{[link]}.
\newline\urlprefix\url{https://www.mathworks.com/matlabcentral/fileexchange/72654-spline-toolbox}

\bibitem{de2005spline}
C.~De~Boor, Spline toolbox for use with MATLAB: User's guide, version 3,
  MathWorks, 2005.

\bibitem{de1978practical}
C.~De~Boor, A practical guide to splines, Vol.~27, springer-verlag New York,
  1978.

\bibitem{collados2019stability}
C.~Collados-Rodriguez, M.~Cheah-Mane, E.~Prieto-Araujo, O.~Gomis-Bellmunt,
  Stability analysis of systems with high vsc penetration: where is the limit?,
  EEE Transactions on Power Delivery 35~(4) (2019) 2021--2031.
\newblock \href {https://doi.org/10.1109/TPWRD.2019.2959541}
  {\path{doi:10.1109/TPWRD.2019.2959541}}.

\bibitem{botchkarev2018performance}
A.~Botchkarev, Performance metrics (error measures) in machine learning
  regression, forecasting and prognostics: Properties and typology, arXiv
  preprint arXiv:1809.03006 (2018).

\end{thebibliography}





% \end{thebibliography}
\end{document}